\documentclass{article}
\usepackage{graphicx}  
\usepackage{amsmath}   
\usepackage{amssymb}   
\usepackage{bm} 
\usepackage{dcolumn}
\usepackage{color}
\usepackage{mathrsfs}
\usepackage{amsfonts}
\usepackage{varioref}
\RequirePackage[colorlinks,citecolor=blue,urlcolor=magenta,linkcolor=blue]{hyperref}
\def\p{\partial}
\def\e{\epsilon}
\def\f{\frac}

\def\be{\begin{equation}}
\def\ee{\end{equation}}

\labelformat{section}{Section #1} 
\labelformat{subsection}{Section #1} 
\labelformat{subsubsection}{Section #1}
\labelformat{subsubsubsection}{Section #1}
\labelformat{equation}{Eq.~(#1)} 
\labelformat{figure}{Fig.~#1} 
\labelformat{subfigure}{Fig.~\thefigure#1} 
\labelformat{table}{Tab.~#1} 
\labelformat{appendix}{Appendix #1}

\title{\bf Massless minimal quantum scalar field  with an asymmetric self interaction in  de Sitter spacetime}
\author{$^{1,2}$Sourav Bhattacharya\footnote{sbhatta.physics@jadavpuruniversity.in}\\
\small{$^1$Relativity and Cosmology Research Centre, Department of Physics, Jadavpur University, Kolkata 700 032, India}\\ 
\small{$^2$Department of Physics, Indian Institute of Technology Ropar, Rupnagar, Punjab 140 001, India\footnote{On lien}}\\}
\begin{document}
\maketitle
\begin{abstract}
\noindent
Massless minimally coupled quantum scalar  field with an asymmetric self interaction, $V(\phi)=\lambda \phi^4/4!+ \beta \phi^3/3! $ (with $\lambda >0$)  is considered in the $(3+1)$-dimensional inflationary de Sitter spacetime. This potential is bounded from below irrespective of the sign of $\beta$. Earlier computations mostly considered the quartic part. Our chief motivation behind this study is to assess the vacuum expectation values of $V(\phi)$ and $\phi$, both of which can be   negative, and hence may lead to some screening of the inflationary cosmological constant value.  First  using the in-in formalism, the renormalised quantum correction to the cubic potential  appearing in the energy-momentum tensor  is computed at two loop, which is the leading order  in this case.  The quantum correction to the kinetic term at two loop are subleading compared to the above result at late cosmological times. Next, using  some of these results we compute the renormalised vacuum expectation value of  $ \phi$, by computing the tadpoles at ${\cal O}(\beta)$ and ${\cal O}(\lambda \beta)$. Due to the appearance of the de Sitter isometry breaking logarithms, the tadpoles cannot be completely renormalised away in this case, unlike the flat spacetime. All these results, as expected, show secularly growing logarithms at late cosmological times.   We next use a recently proposed renormalisation group inspired formalism  to resum perturbative secular effects, to compute a  non-perturbative $\langle \phi \rangle$ at late cosmological times.  $\langle \phi \rangle$ turns out to be approximately one order of magnitude less compared to the position of the classical minima $\phi=-3\beta/\lambda$ of  $V(\phi)$.  Estimation on the possible screening of the inflationary cosmological constant due to this $\langle \phi \rangle$ is also presented.  
\end{abstract}
\vskip .5cm


\newpage
\tableofcontents

\section{Introduction}\label{intro}
The cosmic inflation is a stage of the early universe characterised  by a very rapid accelerated expansion, proposed to solve the famous puzzles associated with the early universe, namely, the horizon problem, the flatness problem and the scarcity of defects like magnetic monopoles,
e.g.~\cite{Weinberg-Book} and references therein.  During this epoch,  quantum fluctuations generate the cosmological correlation functions as well as the seed for the primordial density fluctuations. Such predictions are found to be in good agreement with observation so far.

A massless virtual particle created during this period may last practically forever, e.g.~\cite{Tsamis:2005hd},  
save conformally invariant field theories if they are initially prepared in the conformal vacuum, for in this case there is no particle creation~\cite{Birrell:1982ix}. Thus for particles like massless minimal scalar or gravitons, the quantum effects are expected to be amplified at late cosmological times, owing to the long wavelength super-Hubble modes. Accordingly, any physical process containing propagators of such fields as internal lines are expected to show some kind of infrared divergence at late times,  see~\cite{Floratos:1987ek} and references therein for the earliest of such works. The late time divergence is given by the logarithm of the scale factor, known as the secular effect. These large logarithms indicate breaking of the de Sitter symmetry. We refer our reader to~\cite{Tanaka:2013caa} for a vast review and an exhaustive list of references on this phenomenon.

During inflation, the cosmological constant $\Lambda$ was large, compared to its current observed  tiny value. One of the most important challenges of modern physics is to give a clear mechanism on how exactly the inflation ended into the radiation dominated era, so that the current tiny $\Lambda$ value was reached.   One possible way to address this cosmological constant problem is the screening of the inflationary $\Lambda$ by the aforementioned secular effect~\cite{Tsamis:1992sx}. We also refer our reader to e.g.~\cite{Ringeval:2010hf, Dadhich:2011gx, Padmanabhan:2013hqa, Alberte:2016izw, Appleby:2018yci} and references therein for alternative approaches to address the cosmological constant problem.

If we consider a massless minimal scalar field theory with a quartic self interaction in the inflationary de Sitter spacetime, the secular logarithms can actually be resummed, as was first pointed out in~\cite{Starobinsky, Starobinsky:1994bd} by formulating the late time theory stochastically, showing at late times the de Sitter symmetry is retained, due to dynamical generation of mass. See also  e.g.~\cite{Finelli:2008zg, Vennin:2015hra, Markkanen:2019kpv, Markkanen:2020bfc} and references therein for recent developments on stochastic cosmic inflation.  We further refer our reader to~\cite{Moreau:2018ena,Moreau:2018lmz, Gautier:2015pca, Serreau:2013eoa, Serreau:2013koa, Serreau:2013psa, Ferreira:2017ogo, Burgess:2009bs, Burgess:2015ajz, Youssef:2013by, Baumgart:2019clc, Kamenshchik:2020yyn, Kamenshchik:2021tjh, Moss:2016uix} and references therein for various field theoretic resummation techniques in the same context, such as the non-perturbative renormalisation group, large $N$ scalars, the Schwinger-Dyson resummation etc. We also refer our reader to~\cite{Chu:2015ila} for discussion on dilaton and to~\cite{Kitamoto:2011yx, Kitamoto:2018dek, Miao:2021gic}
on non-linear sigma model.   However, whether the gravitons may generate non-perturbative large late time de Sitter breaking backreaction remains as an open issue till date~\cite{Miao:2021gic, Woodard:2014jba} and references therein.  Note  that even though a massless minimal scalar with a quartic self interaction does not break the de Sitter invariance at late times, the loop effects can bring in interesting testable predictions about the inflationary power spectra, e.g.~\cite{Markkanen:2019kpv}. Also, it is always an  important task to precisely estimate the  backreaction any theory creates on $\Lambda$, even if the de Sitter invariance is retained at late times.

Various one and two loop  field theory in curved space computations for a massless minimal scalar in the de Sitter spacetime can be seen in e.g.~\cite{Onemli:2002hr, Brunier:2004sb, Kahya:2009sz, Boyanovsky:2012qs, Onemli:2015pma, Karakaya:2017evp,  Ali:2020gij, Prokopec:2003tm,  Miao:2006pn, Prokopec:2007ak, Liao:2018sci, Miao:2020zeh, Glavan:2019uni} and references therein. See~\cite{Giddings:2010nc, Leonard:2014zua, Park:2015kua, Frob:2016fcr, Frob:2017smg, Frob:2017lnt, Boran:2017fsx, Boran:2017cfj, Miao:2018bol, Ferrero:2021lhd, Tan:2021ibs, Glavan:2021adm} and references  therein for discussions in the context of one loop perturbative quantum gravity.  We refer our reader to~\cite{Akhmedov:2013xka, Akhmedov:2014doa, Akhmedov:2015xwa, Akhmedov:2019cfd, Kaplanek:2020iay, Burgess:2018sou} for   analogous secular effects in other contexts like near horizon geometry, background fields etc. We refer our reader to~\cite{Hu:2018nxy} and references therein for  an extensive review and list of references on  infrared effects in the inflationary spacetime and its possible observational consequences. 
	\begin{figure}[h!]
		\includegraphics[height=4.5cm]{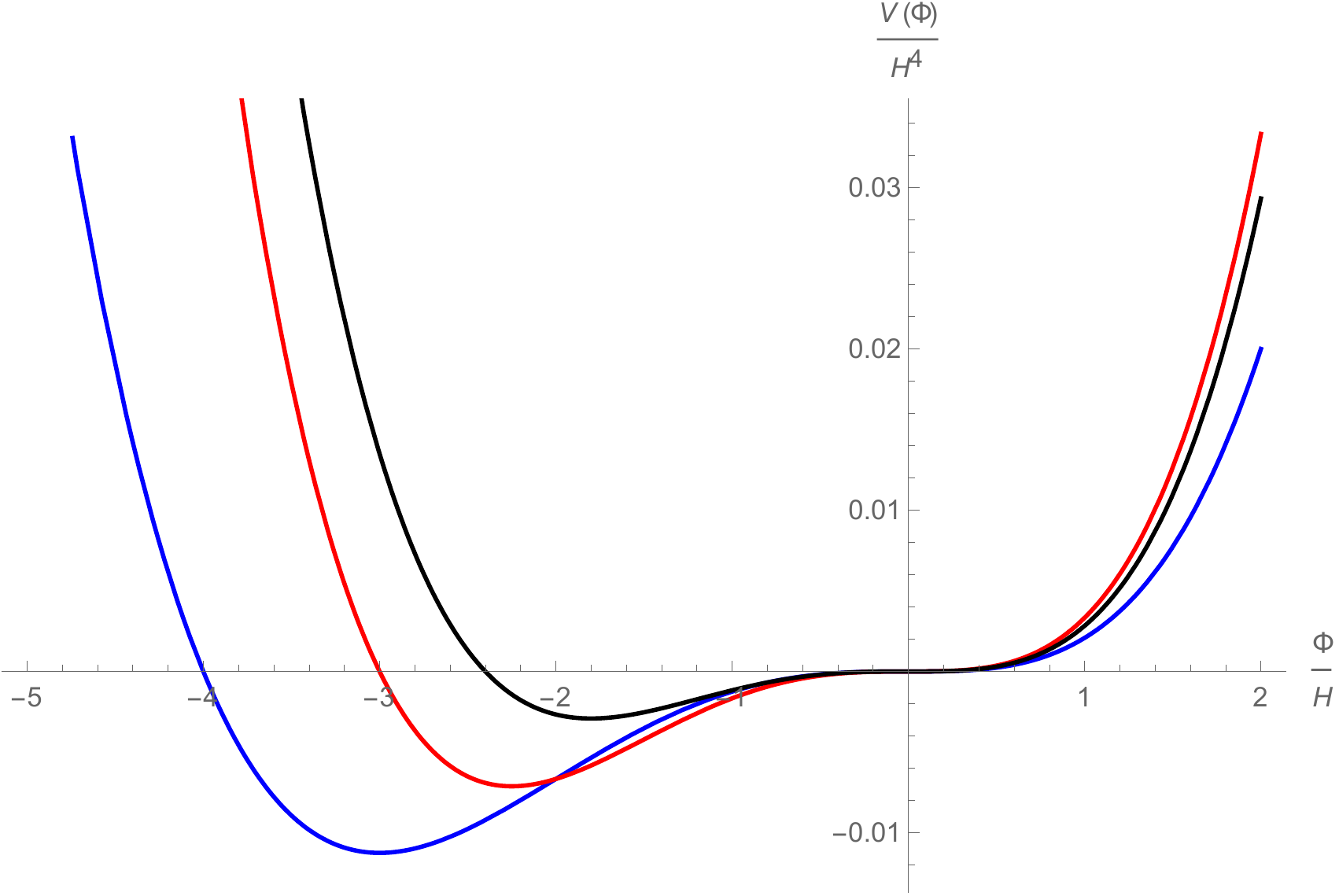}\centering \hskip 2cm
		\caption{ \footnotesize A typical plot of  $V(\phi)=\lambda \phi^4/4!+\beta \phi^3/3!$ with respect to the field $\phi$, both being scaled respectively by $H^4$ and $H$, where $H$ is the Hubble constant. The blue, red and black curves respectively correspond to $\lambda$ and $\beta/H$ values : $0.01,~0.01$; $ 0.02,~0.015$; and  $0.02,~0.012$. The  minimum of the potential is located at $\phi=-3\beta/\lambda$.    The system is assumed to be located at the flat plateau around $\phi \sim 0$ initially ($t\to0$ in \ref{l0-})}.
		\label{f1}
	\end{figure}

 In this work we wish to investigate a massless and minimal quantum scalar field theory in the inflationary de Sitter spacetime with a bounded from below self interaction, $V(\phi)=\lambda \phi^4/4!+\beta \phi^3/3!$ $(\lambda >0)$, \ref{f1}.     The quartic case  is much well studied in de Sitter, whereas the cubic case is addressed in some earlier works too. For example, some non-perturbative results with $\lambda=0$ for a massive scalar field, derived via the stochastic formalism can be seen in the pioneering work~\cite{Starobinsky}. See~\cite{Burgess:2009bs} for a renormalisation group discussion on the cubic self interaction in de Sitter. We also refer our reader to~\cite{Martins:2020oxv} for a discussion on cubic self interaction in the presence of non-minimal coupling and fermions. To the best of our knowledge however, the quartic plus cubic self interactions  has not been considered as a single potential yet. Our motivation is as follows.  From \ref{f1} we may expect that the late time vacuum expectation value of $V(\phi)$ may be negative. Apart from this, there will be vacuum expectation value of $\phi$ owing to the tadpoles, which cannot be completely renormalised away due to the appearance of the secular logarithms. Even though we do expect that these  loop effects will be resummable and the de Sitter symmetry will be retained at late times, it is nevertheless an important task to estimate their precise backreaction on the inflationary $\Lambda$.   For $\beta=0$, the potential's contribution will be positive whereas that of the field's will be vanishing. On the other hand for $\lambda=0$, the potential becomes unbounded from below and would lead the system to eternal instability. Note also that an asymmetric potential might lead to interesting predictions in the inflationary power spectra, e.g.~\cite{Markkanen:2020bfc}.

 The rest of the paper is organised as follows. In the next section, we briefly review the basic setup necessary for our computations. In \ref{s3}, we compute the ${\cal O}(\beta^2)$ two loop correction to the cubic potential term appearing in the energy-momentum tensor and the result is compared with that of the quartic self interaction.  Using some of these results, in \ref{tadpole}, we compute the one and two loop (${\cal O}(\beta)$, ${\cal O}(\lambda \beta)$) vacuum expectation values of the field operator. We find a resummed expression for the same in \ref{resum}, using the formalism recently proposed in~\cite{Kamenshchik:2020yyn, Kamenshchik:2021tjh}. Clearly, this value  cannot be expected to correspond to the minima of the potential in \ref{f1}, as the shape of the potential would certainly change due the quantum effects. Finally, we conclude in \ref{disc} with an estimate of the backreaction created by this  non-perturbative tadpole contribution to the inflationary $\Lambda$, along with some possible future directions. At initial times, the system is assumed to be located around $\phi \sim 0$ in \ref{f1}, such that the perturbation over a free massless minimal scalar is well defined at that time. The details of the calculations are provided in  appendices. The coordinate space methods developed in \cite{Onemli:2002hr, Brunier:2004sb} will be useful for our purpose.    We shall work with mostly positive signature of the metric in $d=4-\e$ ($\e=0^+$) spacetime dimensions and will set $c=1=\hbar$ throughout. Also in the following, $\ln^n a$ will always stand for $(\ln a)^n$.

\section{The basic setup}\label{setup}
We shall briefly review below the basic setup we shall be using for our computations,  chiefly be based upon the discussion of~\cite{Onemli:2002hr, Brunier:2004sb} and references therein. The metric of the cosmological or inflationary de Sitter spacetime reads,
\be
ds^2 = -dt^2 + e^{2Ht}d\vec{x}\cdot  d\vec{x}
\label{l0-}
\ee
where the Hubble constant $H=\sqrt{\Lambda/3}$, with $\Lambda$ being the positive cosmological constant. The above metric in a conformally flat form reads,
\begin{eqnarray}
ds^2 = a^2(\eta) \left[-d\eta^2 + d\vec{x}\cdot  d\vec{x}\right],
\label{l0}
\end{eqnarray}
where $a(\eta):= - 1/H\eta $ and the conformal time $\eta = - e^{-H t}/H$. We shall  set our initial conditions at $t=0$ or at $\eta=-1/H$. The spacetime dimension will be taken to be $d=4-\e$ ($\e >0$).

The Lagrangian density for a Hermitian, massless and minimally coupled scalar field we shall be working on reads
\begin{eqnarray}
{\cal L}= -\frac12 (\nabla_{\mu} \phi)(\nabla^{\mu} \phi) - \frac{\lambda}{4!}\phi^4- \frac{\beta}{3!}\phi^3 + \Delta{\cal L}
\label{l1}
\end{eqnarray}
where $\lambda>0$ whereas $\beta$ can be either positive or negative.  The counterterm Lagrangian density $\Delta {\cal L}$ reads
\begin{eqnarray}
\Delta {\cal L}= -\frac12 \delta m^2 \phi^2 + \delta \xi \left(R- d(d-1)H^2 \right)\phi^2  -\frac{\delta \Lambda}{8\pi G} -\frac12 \delta Z  (\nabla_{\mu} \phi)(\nabla^{\mu} \phi) - \frac{\delta \lambda}{4!}\phi^4-\frac{\delta \beta }{3!} \phi^3 -\alpha \phi
\label{l2}
\end{eqnarray}
$\delta \xi$ is the conformal counterterm will be required later.  The energy-momentum tensor corresponding to \ref{l1}, \ref{l2}, is given by
\begin{eqnarray}
T_{\mu \nu} (x)&&
= (1+\delta Z) \left(\delta_{\mu}{}^{\alpha}\delta_{\nu}{}^{\beta} -\frac12 \eta^{\alpha \beta} \eta_{\mu \nu} \right) (\partial_{\alpha} \phi)(\partial_{\beta} \phi) - \frac12 \delta m^2 \phi^2 g_{\mu \nu} -\frac{\lambda+ \delta \lambda}{4!}  \phi^4g_{\mu \nu}-\frac{\beta+ \delta \beta}{3!}\phi^3 g_{\mu \nu} \nonumber\\&& - \alpha \phi g_{\mu \nu}
 -\frac{\delta \Lambda}{8\pi G} g_{\mu \nu} -2 \delta \xi\left[ (d-1)H^2 g_{\mu \nu} + \left(\eta_{\mu\nu} \eta^{\rho \lambda} - \delta_{\mu}{}^{\rho} \delta_{\nu}{}^{\lambda}\right)  \left(\partial_{\rho}\partial_{\lambda}- \Gamma^{\alpha}_{\rho \lambda}\partial_{\alpha}\right) \right]\phi^2 
\label{l3}
\end{eqnarray}
In this paper, we shall chiefly be focussed on the loop corrections to the vacuum expectation values of the cubic potential  and the field operator $\phi$. The same for the kinetic and the quartic self interaction terms can be seen in~\cite{Onemli:2002hr}. We shall not compute the loop correction to the kinetic term  here, as the late time contribution of the same due to the derivatives, will be subleading compared to that of the potential.

The mode function for a free massless and minimally coupled scalar, $\Box \phi=0$ in \ref{l0}, behaving as  positive frequency in the asymptotic past  reads, e.g.~\cite{Tsamis:2005hd},
\begin{eqnarray}
\phi_{\vec k}(x)= \frac{H \left(1+ik\eta \right)}{\sqrt{2k^3}} e^{-i(k\eta-\vec{k}\cdot \vec{x})}
\label{l3'}
\end{eqnarray}
where ${\vec k}$ ($-\infty < k^i < \infty$)  is the spatial momentum associated with the mode, with $k =\sqrt{k_i k_i}$. At the initial time, $\eta =-1/H$, the modes are sub-horizon ($k/H \gg 1$), giving 
$$\phi_{\vec k}(\eta \to -1/H,{\vec x})\to  \frac{H}{\sqrt{2k}} e^{-i(k\eta-\vec{k}\cdot \vec{x})}$$
which is similar to that of the flat spacetime. On this initial hypersurface it is easy to see using the Klein-Gordon inner product that the above mode functions are $\delta$-function normalisable. One then quantises the free field and the associated vacuum is known as the Bunch-Davies vacuum, say $|0\rangle$. It is however also easy to see that the Bunch-Davies mode functions \ref{l3'} are not normalisable on hypersurfaces lying in the future of $\eta =-1/H$. Consequently,  the free Wightman functions or the propagators for  such a field do not satisfy de Sitter invariance, analogous to that of the gravitons in de Sitter, e.g.~\cite{Tan:2021ibs}.

Due to the particle creation effects, the initial Bunch-Davies vacuum is not stable and will evolve into a qualitatively different `out' vacuum.  Due to this reason, the standard  `in-out' S-matrix formalism of  quantum field theory requiring a unique vacuum fails in this case. Accordingly, in order to compute the expectation value of any operator, we need the in-in or the closed time path or the Schwinger-Keldysh formalism, e.g.~\cite{Calzetta:2008iqa, Chou, Calzetta1, Calzetta2, Weinberg, Adshead:2009cb} and references therein. Also note in particular that any matrix element between the in and out vacua does not yield any causal dynamics, the out vacuum being defined in the asymptotic  future. Let us now have a very quick review of the in-in formalism.

The time ordered functional integral representation of the standard in-out matrix elements in terms of the field basis reads,
\begin{eqnarray}
\langle \phi | T (A[\phi]) | \psi \rangle = \int {\cal D} \phi e^{i \int_{t_i}^{t_f} \sqrt{-g} d^dx {\cal L}[\phi] } \Phi^{\star}[\phi(t_f)] A[\phi] \Psi[\phi(t_i)]
\label{ii1}
\end{eqnarray}
where $A[\phi]$ is an observable, and any derivative is taken outside the time ordering.  $\Phi$ and $\Psi$ are the wave functionals with respect to the field kets $|\phi\rangle$ and $|\psi\rangle$ respectively.  However, due to the aforementioned reason, computation of the above matrix elements in a dynamical background such as \ref{l0} is not very meaningful. Hence in such a background one uses the in-in functional representation as follows. We may use \ref{ii1} to write to write for a functional $B[\phi]$,
\begin{eqnarray}
\langle \psi | \overline{T} (B[\phi]) | \phi \rangle =\left(\langle \phi | T (B[\phi])^{\dagger} | \psi \rangle\right)^{\star}= \int {\cal D} \phi e^{-i \int_{t_i}^{t_f} \sqrt{-g} d^dx {\cal L}[\phi] } \Phi[\phi(t_i)] B[\phi] \Psi^{\star}[\phi(t_f)]
\label{ii1'}
\end{eqnarray}
where $\overline{T}$ stands for anti-time ordering.  The completeness relation on the final hypersurface at $t=t_f$ reads,
$$ \int {\cal D} \phi \Phi[\phi_-(t_f)] \Phi^{\star}[\phi_+(t_f)]=\delta(\phi_+(t_f)-\phi_-(t_f)) $$
 Multiplying now \ref{ii1} with \ref{ii1'} and using the above, we have
\begin{eqnarray}
\langle \psi | \overline{T} (B[\phi]) T (A[\phi]) | \psi \rangle = \int {\cal D} \phi_+  {\cal D} \phi_-  \delta(\phi_+(t_f)- \phi_-(t_f))  e^{i \int_{t_i}^{t_f} \sqrt{-g} d^dx \left({\cal L}[\phi_+]-{\cal L}[\phi_-] \right)} \Psi^{\star}[\phi_-(t_i)] B[\phi_-] A[\phi_+] \Psi[\phi_+(t_i)] 
\label{ii2}
\end{eqnarray}
Note that $\phi_+$ and $\phi_-$ are two different scalar fields, the former evolves the system forward in time whereas the latter evolves it backward, and they are coincident on the final hypersurface. Both ${\cal L}[\phi_+]$ and ${\cal L}[\phi_-]$  are given by \ref{l1}, with $\phi$ being replaced either by $\phi_+$ or $\phi_-$.  We shall compute expectation values with respect to the initial Bunch-Davies vacuum $|0\rangle$, defined above. 

Some terms relevant for our purpose with $\phi_+$  are given by
\begin{eqnarray}
&& \eta^{\mu \nu } \p_{\mu} \left(a^{d-2} \p_{\nu} i \Delta (x,x') \right) = i \delta^d (x-x') \qquad \qquad \qquad \quad  ({\rm Propagator}) \nonumber \\
&&- i(\lambda+\delta \lambda )a_1^d \delta^d(x_1-x_2) \delta^d(x_2-x_3) \delta^d(x_3-x_4)~\qquad \,\,\,({\rm 4-point ~ vertex~with~counterterm}) \nonumber \\
&&- i(\beta+\delta \beta)  a_1^d \delta^d(x_1-x_2) \delta^d(x_2-x_3) ~~\qquad \qquad  \qquad \quad ({\rm 3-point ~ vertex~with~ counterterm}) \nonumber \\
&& -i \delta m^2 a^d \delta^d(x-x')  \qquad \qquad \qquad \qquad \qquad   \qquad \,\qquad \quad  ({\rm Mass~renormalisation}) \nonumber \\
&&- i\alpha a^d \qquad \qquad  \qquad \qquad \qquad \qquad \qquad   \qquad \quad\,\,\,\, \qquad \quad  ({\rm Tadpole~renormalisation}) 
\label{l5}
\end{eqnarray}
where $a_1 \equiv a(\eta =\eta_1)$. For the field $\phi_-$, the interaction terms appearing above will carry a  `minus' sign in front of them.  The field strength renormalisation $\delta Z$ will not be necessary for our current purpose.

The propagator $i \Delta(x,x')$ in spacetime dimensions $d=4-\epsilon$ can be split into three parts~\cite{Onemli:2002hr, Brunier:2004sb},
\begin{eqnarray}
i\Delta(x,x')= A(x,x') + B(x,x') + C(x,x'),
\label{l6}
\end{eqnarray}
where,
\begin{eqnarray}
&&A(x,x')= \frac{H^{2-\e}\,\Gamma\left(1-\frac{\epsilon}{2}\right)}{4\pi^{2-\frac{\epsilon}{2}}} \frac{1}{y^{1-\frac{\e}{2}}} \nonumber\\
&& B(x,x')= \frac{H^{2-\e}}{(4\pi)^{2-\frac{\e}{2}}} \left[-\frac{2 \Gamma(3-\frac{\e}{2})}{\e} \left(\frac{y}{4} \right)^{\frac{\e}{2}} +   \frac{2}{\e}\frac{\Gamma(3-\e)}{\Gamma(2-\frac{\e}{2})}+  \frac{\Gamma(3-\e)}{\Gamma(2-\frac{\e}{2})} \,\ln (aa')\right] \nonumber\\
&& C(x,x')=\frac{H^{2-\e}}{(4\pi)^{2-\frac{\e}{2}}} \sum_{n=1}^{\infty}\left[  \f{\Gamma(3-\e+n)}{n\,\Gamma(2-\f{\e}{2}+n)} \left(\f{y}{4} \right)^{n} - \f{\Gamma\left(3- \f{\e}{2}+n\right)}{\left(n+ \f{\e}{2}\right)\, \Gamma(n+2) } \left(\f{y}{4} \right)^{n+ \f{\e}{2}}  \right]
\label{l7}
\end{eqnarray}
where $y$ is a biscalar  given by
$$y(x,x') =aa' H^2 \Delta x^2=aa' H^2 \left[-(\eta-\eta')^2+ |{\vec x}-{\vec x'}|^2\right]$$ 
However, $\Delta x^2$ needs to be 
 modified by suitable $i\e$ prescriptions as follows. Due to the necessity of moving both forward and backward in time in \ref{ii2}, we have four propagators in the  in-in formalism : a) $i\Delta_{++} (x,x')$ (the Feynman propagator) b) $i\Delta_{-+} (x,x')$ and  $i\Delta_{+-} (x,x')$ (the positive and negative frequency Wightman functions, both satisfying the homogenous version of the first of \ref{l5}, with ($ i\Delta_{+-} (x,x'))^{\star} = i\Delta_{-+} (x,x')$) and c) $i\Delta_{--} (x,x')$ (the anti-time ordered propagator, with $(i\Delta_{++} (x,x'))^{\star}=i\Delta_{--} (x,x')$). Corresponding to these four propagators, the  biscalar $y(x,x')$ in \ref{l7} respectively takes four expressions 
\begin{eqnarray}
y_{++}&=& aa' H^2 \Delta x_{++}^2 = aa' H^2  \left[ (\vec{x}- \vec{x'})^2 - \left(|\eta -\eta'| -i \e \right)^2\right] = (y_{--})^{\star}\nonumber\\
y_{+-}&=& aa' H^2 \Delta x_{+-}^2 = aa' H^2  \left[ (\vec{x}- \vec{x'})^2 - \left((\eta -\eta') +i \e \right)^2\right]= (y_{-+})^{\star}\nonumber\\
\label{l8}
\end{eqnarray}
We also note that
\be
i\Delta_{++} (x,x') = \theta(\eta-\eta') i\Delta_{-+}(x,x')+\theta(\eta'-\eta) i\Delta_{+-}(x,x')
\label{feyn}
\ee
where,
$$i\Delta_{-+}(x,x')=\langle 0| \phi_-(x)\phi_+(x')|0\rangle,\qquad i\Delta_{+-}(x,x')=\langle 0| \phi_-(x')\phi_+(x)|0\rangle  = i\Delta_{-+}(x',x)$$
Finally, for all of the above propagators, we have from \ref{l7},
\begin{eqnarray}
\lim_{\e \to 0} C(x,x') =0, \qquad  {\rm and} \qquad
\lim_{\e \to 0} B(x,x')= - \frac{H^2}{8\pi^2} \ln \frac{\sqrt{e} H^2 \Delta x^2}{4}
\label{l8'}
\end{eqnarray}
Using the above  it is easy to see from \ref{l6}, \ref{l7} and  \ref{l8} that for {\it all} the four propagators, we have in the coincidence limit
\begin{eqnarray}
i \Delta (x,x )= \frac{H^{2-\e}}{2^{3-\e}\pi^{2-\e/2}} \frac{\Gamma(3-\e)}{\Gamma(2-\frac{\e}{2})} \left(\frac{1}{\e}+ \ln a \right)
\label{l13}
\end{eqnarray}
Note in particular the appearance of the logarithm of the scale factor above, due to the breaking of the de Sitter invariance by a massless minimal scalar. Such logarithms will be responsible  for  the key interesting features of our calculations. 

With these ingredients, we are ready to compute the renormalised vacuum expectation values for the cubic and linear terms of our scalar field theory. 

\section{Loop correction to the cubic potential}\label{s3}

We shall compute below the 2-loop correction to the cubic potential term appearing in \ref{l3} due to the cubic self interaction itself. This is the lowest order quantum correction to it. However before that we would like to very quickly look into some computational tools  of~\cite{Onemli:2002hr} in the context of the quartic self interaction, which we shall need for our purpose. For any operator $A$ appearing below, $\langle 0|A |0\rangle$ will be abbreviated as $\langle A \rangle$.

\subsection{The case of quartic self interaction -- a quick look}\label{lcphi4}
%
	\begin{figure}[h!]
		\includegraphics[height=2.5cm]{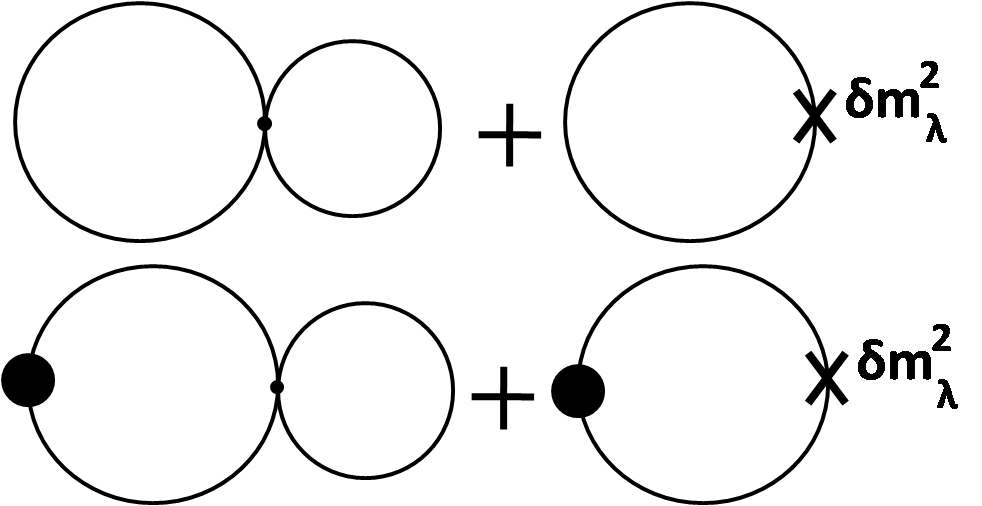}\centering \hskip 2cm
		\caption{\footnotesize Two loop correction to the 
		energy-momentum tensor due to the quartic self interaction, discussed in~\cite{Onemli:2002hr}. The first row corresponds to the vacuum expectation value of the quartic potential itself, whereas the second row corresponds to the correction to the kinetic term. The big blob corresponds to the kinetic vertex containing the derivatives. $\delta m_{\lambda}^2$ is the one loop ${\cal O}(\lambda)$ mass counterterm, \ref{l47'}. }
		\label{f2}
	\end{figure}

Let us first compute the two loop contribution from the $\lambda \phi^4$ term in \ref{l3}, which corresponds to the vacuum expectation value of itself plus the mass  renormalisation counterterm contribution, as shown in the first row of \ref{f2}. We have
\begin{eqnarray}
\langle T_{\mu\nu}^{(\phi^4)}\rangle = -\left[ \frac{\lambda}{8}(i\Delta (x,x))^2+ \frac{\delta m^2}{2} i\Delta (x,x)\right] \, g_{\mu\nu}
\label{l12}
\end{eqnarray}
The mass renormalisation counterterm $\delta m^2$ is found by computing 
\begin{eqnarray}
-i \left[\frac{\lambda}{2} i\Delta_{++} (x,x) +\delta m^2 \right] a^d \delta^{d}(x-x'),
\label{l14}
\end{eqnarray}
if the interaction vertex contains $\phi_+$ field.  Using \ref{l13}, we have 
\begin{eqnarray}
\delta m^2_{\lambda} = -\frac{\lambda H^{2-\e}} {(4 \pi)^{2-\e/2}}\frac{\Gamma(3-\e)}{\Gamma(2-\e/2)\e} 
\label{l47'}
\end{eqnarray} 
$\delta m^2_{\lambda}$ remains the same if the interaction vertex consists of $\phi_-$. Substituting now \ref{l13}, \ref{l47'} into \ref{l12}, we have 
\begin{eqnarray}
\langle T_{\mu\nu}^{(\phi^4)}\rangle = \left[ \frac{\lambda H^{4-2\e}}{2^{7-2\e} \pi^{4-\e}}\frac{\Gamma^2(2-\e)}{\Gamma^2(1-\e/2)\e^2} - \frac{\lambda H^4}{2^7 \pi^4}\ln^2 a
\right]  g_{\mu\nu}
\label{l12'}
\end{eqnarray}
The first term is a constant and hence needs to be absorbed via a cosmological constant counterterm $\delta \Lambda$ appearing in \ref{l3}. The remaining term containing $\ln^2 a$ is ultraviolet finite but grows monotonically with time, indicating breakdown of the perturbation theory at sufficiently late cosmological times, requiring the need of a non-perturbative treatment for such secular effect. 

Let us now see very briefly the two loop correction due to the quartic self interaction to the kinetic term. The free kinetic term's contribution in \ref{l3} reads,
\begin{eqnarray}
\langle T_{\mu\nu}^{{\rm kin.}}\rangle = \Omega_{\mu\nu}^{\alpha\beta} \partial_{\alpha} \partial'_{\beta} \langle \phi_+(x)  \phi_+(x')\rangle\big\vert_{x\to x'}=\Omega_{\mu\nu}^{\alpha\beta} \partial_{\alpha} \partial'_{\beta} i\Delta_{++}(x,x')\big\vert_{x\to x'}
\label{k1}
\end{eqnarray}
where we have abbreviated $\Omega_{\mu\nu}^{\alpha\beta}=\left(\delta_{\mu}{}^{\alpha}\delta_{\nu}{}^{\beta} - \eta^{\alpha \beta} \eta_{\mu \nu}/2 \right)$. We now substitute \ref{l6}, \ref{l7}, \ref{l8} into the above equation. Only the $n=1$ term contributes in the limits $x\to x'$ and $\e \to 0$, giving
\begin{eqnarray}
\langle T_{\mu\nu}^{{\rm kin.}}\rangle = \frac{3H^4}{2^6 \pi^2} g_{\mu\nu} 
\label{k2}
\end{eqnarray}
which can be absorbed in the cosmological counterterm $\delta \Lambda$ in \ref{l3}. The ${\cal O}(\lambda)$ two-loop correction to the kinetic term reads (second row of \ref{f2})
\begin{eqnarray}
&&\langle T_{\mu\nu}^{{\rm kin.},\lambda}\rangle = \Omega_{\mu\nu}^{\alpha\beta} \int a'^dd^dx' \left(\p_{\alpha} i\Delta_{++}(x,x')\p_{\beta} i\Delta_{++}(x,x') -\p_{\alpha} i\Delta_{+-}(x,x')\p_{\beta} i\Delta_{+-}(x,x')\right)\left(-\frac{i\lambda}{2} i\Delta(x',x')-i\delta m^2\right)\nonumber\\
&&= -\frac{i\lambda H^{2-\e}}{2^{3-\e} \pi^{2-\e/2}}\frac{\Gamma(2-\e)}{\Gamma(1-\e/2)}\Omega_{\mu\nu}^{\alpha\beta} \int a'^dd^dx' \ln a' \left(\p_{\alpha} i\Delta_{++}(x,x')\p_{\beta} i\Delta_{++}(x,x') -\p_{\alpha} i\Delta_{+-}(x,x')\p_{\beta} i\Delta_{+-}(x,x')\right)
\label{k3}
\end{eqnarray}
where in the last line we have used \ref{l13} and \ref{l47'}. Substituting \ref{l6}, \ref{l7} and \ref{l8} into the above, we have
\begin{eqnarray}
&&\p_{\alpha} i \Delta_{++}(x,x')\p_{\beta} i \Delta_{++}(x,x')-\p_{\alpha} i \Delta_{+-}(x,x')\p_{\beta} i \Delta_{+-}(x,x') \nonumber\\
&&= \frac{H^{6-2\e} \Gamma^2(2-\e/2)a^2}{2^4 \pi^{4-\e}}\left[ 4 a'^2 H^2 \Delta x^{\alpha} \Delta x^{\beta}\left(\frac{1}{y_{++}^{4-\e} }- \frac{1}{y_{+-}^{4-\e} }\right)  + (4-\e) a'^2 H^2 \Delta x^{\alpha} \Delta x^{\beta}\left(\frac{1}{y_{++}^{3-\e} }- \frac{1}{y_{+-}^{3-\e} }\right)\right.\nonumber\\ &&\left. + \left(1-\frac{\e}{2}\right) a'^2 H^2 \Delta x^{\alpha} \Delta x^{\beta}\left(\frac{1}{y_{++}^{2} }- \frac{1}{y_{+-}^{2} }\right)+ 2 a' H \left(\Delta x^{\alpha} \delta^{\beta}{}_0+ \Delta x^{\beta}  \delta^{\alpha}{}_0\right)\left(\frac{1}{y_{++}^{3-\e} }- \frac{1}{y_{+-}^{3-\e} }\right)
\right.\nonumber\\ &&\left.
+  a' H \left(\Delta x^{\alpha} \delta^{\beta}{}_0+ \Delta x^{\beta}  \delta^{\alpha}{}_0\right)\left(\frac{1}{y_{++}^{2} }- \frac{1}{y_{+-}^{2} }\right)+ \delta^{\alpha}{}_0\delta^{\beta}{}_0 \left(\frac{1}{y_{++}^{2-\e} }- \frac{1}{y_{+-}^{2-\e} }\right)\right],
\label{l21}
\end{eqnarray}
which splits \ref{k3} into six sub-integrals. For our future purpose, we have outlined the evaluation of the first of them in \ref{A}. We have the unrenormalised contribution for the kinetic term,
\begin{eqnarray}
\langle T_{00}\rangle^{\rm K}_{\phi^4}
&&=\frac{\lambda a^2 H^4}{2^6 \pi^4}\left[\left( -\frac{\zeta}{2\e}-\frac12 \ln \frac{2\mu}{H}+\frac76\right) \ln a + \frac12\left(\frac{\zeta}{\e}+ \ln \frac{2\mu}{H}+\frac{\pi^2}{6}-\frac{22}{9} \right)  +\frac{2}{9a^3} - \sum_{n=1}^{\infty}\frac{(n+2)\,a^{-n-1}}{2(n+1)^2}\right] 
\nonumber\\ 
\langle T_{ij}\rangle^{\rm K}_{\phi^4}&&= \frac{\lambda a^2 H^4 }{2^6 \pi^4} \left[\left(\frac{\zeta}{2\e}+\frac12 \ln \frac{2\mu}{H} -\frac32 \right) \ln a-\frac13 \left(\frac{\zeta}{\e}+\ln \frac{2\mu}{H}+\frac{\pi^2}{4}-\frac83 \right) -\sum_{n=1}^{\infty} \frac{n^2-4}{6(n+1)^2} a^{-n-1}\right]\delta_{ij}
\label{l41}
\end{eqnarray} 
where the constant $\zeta$ is defined in \ref{l36}. Comparison of the above result with that of the potential term \ref{l12'}, shows that at late times the latter gives dominant contribution. This is due to the fact that from each internal line in the Feynman diagram, we expect a secular logarithm at late times. Thus even though \ref{k3} contains three internal lines, the two derivatives eventually yield a secular growth only as  $\ln a $ at late times. Combining the above with \ref{l12'}, the full two loop unrenormalised result for the quartic self interaction follows.

In order to renormalise this result, one needs  both cosmological constant as well as the conformal counterterms, as of \ref{l3}. The contribution of the conformal counterterm in \ref{l3}, equals
\begin{eqnarray}
\langle T_{\mu\nu}^{\delta \xi}\rangle&&=-2\delta \xi \left[ (d-1)H^2 g_{\mu\nu}+ \left(\eta_{\mu\nu} \eta^{\rho\lambda} -\delta_{\mu}{}^{\rho} \delta_{\nu}{}^{\lambda}\right)\left(\p_{\rho} \p_{\lambda} -\Gamma^{\alpha}_{\rho\lambda} \p_{\alpha} \right)\right] i \Delta(x,x) \nonumber\\
&&= \frac{\delta \xi a^2 H^{4-\e}}{2^{1-\e}\pi^{2-\e/2}} \frac{\Gamma(2-\e)}{\Gamma\left(1-\frac{\e}{2}\right)}\left[-\eta_{\mu\nu}  \left(\frac{3}{\e} +(3-\e)\ln a\right)+  \left((3-\e)\eta_{\mu\nu} -\delta_{\mu}{}^0 \delta_{\nu}{}^0\right)\right]
\label{l42}
\end{eqnarray} 
A careful choice of $\delta \Lambda$ and $\delta\xi$ from \ref{l3} can be used to absorb the divergences, the finite constants as well as the arbitrary mass scale $\mu$ dependent  terms in \ref{l12'} and \ref{l41}. Defining then the energy density and pressure respectively as $\rho = -\langle T_0{}^0\rangle$ and $ P\delta_{ij}=\langle T_i{}^j\rangle $, one obtains the fully renormalised and dynamical result
\begin{eqnarray}
\rho_{\lambda,\,{\rm ren.}}= \frac{\lambda H^4 }{2^7 \pi^4}\left[ \ln^2 a  +\frac{4}{9a^3} - \sum_{n=1}^{\infty}\frac{(n+2)\,a^{-n-1}}{(n+1)^2}\right], \qquad 
P_{\lambda,\,{\rm ren.}}=-\frac{\lambda H^4 }{2^7 \pi^4} \left[ \ln^2 a +\frac23 \ln a+ \sum_{n=1}^{\infty} \frac{(n^2-4)a^{-n-1}}{3(n+1)^2} \right]\nonumber\\ 
\label{l47}
\end{eqnarray} 
The leading late time secular dominance is given by $\sim\lambda \ln^2 a$ (with $\rho>0$ and $P<0$). At late times the equation of state is given by, $(\rho_{\lambda,\,{\rm ren.}}+P_{\lambda,\,{\rm ren.}}) \sim - \lambda\ln a <0$, indicating violation of the dominant energy condition. This in turn, if we consider backreaction, implies positivity of the rate of change of the Hubble rate, $\dot{H} >0$, and hence cosmic superacceleration as pointed out 
in~\cite{Onemli:2002hr},    as long as $\lambda Ht \lesssim {\cal O}(1) $.  However at late times such effects are resummable~\cite{Starobinsky:1994bd} (also e.g.~\cite{Moreau:2018lmz} for discussion with an $O(N)$ model), so that de Sitter symmetry is retained. This is due to the dynamical generation of mass of the scalar at late times. A massive scalar, no matter how small its mass is, does not lead to any de Sitter symmetry breaking. 

As we mentioned earlier, a positive definite potential like  $ \lambda \phi^4/4!$ will always give a positive shift in the cosmological constant at late times, as can be readily seen using, e.g. the infrared effective  stochastic techniques~\cite{Starobinsky:1994bd, Finelli:2008zg}. However, since the potential we are considering, \ref{f1}, is not positive definite, we may expect such potentials would  give rise to a negative vacuum expectation value, for both $T_{\mu\nu}$ and $\phi$, which we wish to compute below.

\subsection{Two loop correction to the cubic potential}
\label{pot}

\begin{figure}[h!]
		\includegraphics[height=2.5cm]{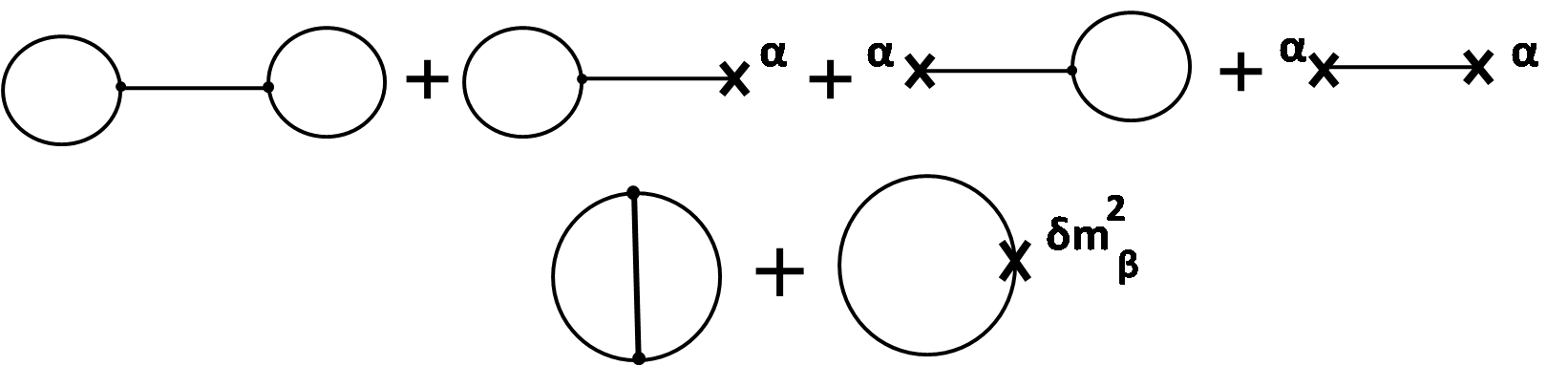}\centering \hskip 2cm
		\caption{ \footnotesize Feynman diagrams corresponding to \ref{l48}. $\alpha$ and $\delta m_{\beta}^2$ respectively correspond to the one loop tadpole  and the one loop mass renormalisation counterterms for the cubic self interaction. }
		\label{f3}
	\end{figure}

There are two two-loop processes we need to consider here, as shown in the \ref{f3}.
The contribution equals (recalling that all propagators yield the same coincidence limit, \ref{l13}),
\begin{eqnarray}
&&\langle T_{\mu\nu}^{\phi^3}\rangle=\frac{i\beta^2\,g_{\mu\nu}}{4} \int d^d x' a'^d\, i \Delta(x,x) i \Delta(x',x') \left[ i\Delta_{++}(x,x') -i\Delta_{+-}(x,x')\right]\nonumber\\&&+\frac{i\beta^2\,g_{\mu\nu}}{6} \int d^d x' a'^d\,\left[ (i \Delta_{++}(x,x'))^3 -(i\Delta_{+-}(x,x'))^3\right]  +\, {\rm contributions~~from~~one~~loop~~ counterterms} \qquad 
\label{l48}
\end{eqnarray} 
	\begin{figure}[h!]
		\includegraphics[height=2.5cm]{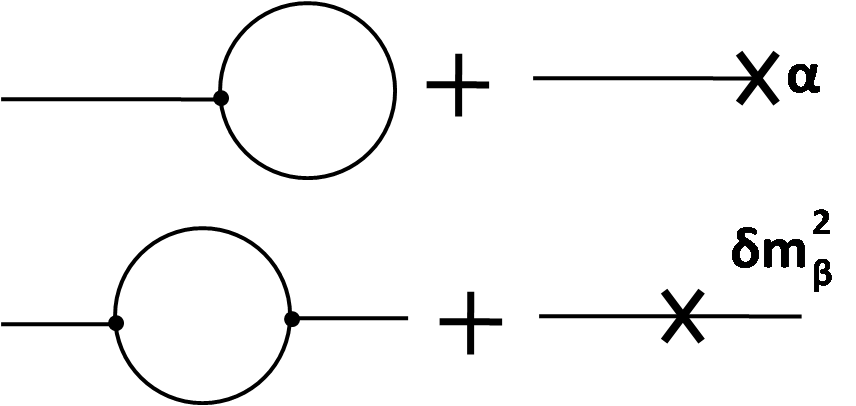}\centering \hskip 2cm
		\caption{\footnotesize Amputated diagrams for computing the one loop tadpole  and the one loop mass renormalisation counterterms for the cubic self interaction.}
		\label{f4}
	\end{figure}
The contribution from the counterterms comes from the tadpoles (for the first integral) and the mass renormalisation (for the second integral), computed in \ref{B}. Let us first evaluate the first integral along with the tadpole counterterm contributions. Using \ref{l13}, we have for the first of \ref{l48}
\begin{eqnarray}
&& \frac{i\beta^2 H^{4-2\e}}{2^{8-2\e} \pi^{4-\e}} \frac{\Gamma^2(3-\e)}{\Gamma^2(2-\e/2)}\left(\frac{1}{\e}+\ln a \right) \int d^d x' a'^d\, \left(\frac{1}{\e}+\ln a' \right)  \left[ i\Delta_{++}(x,x') -i\Delta_{+-}(x,x')\right]
 \nonumber\\&&+ \frac{i\alpha \beta}{2} \frac{H^{2-\e}}{2^{3-\e} \pi^{2-\e/2}} \frac{\Gamma(3-\e)}{\Gamma(2-\e/2)} \int d^d x' a'^d \left(\frac{1}{\e} +\ln a\right) \left[ i\Delta_{++}(x,x') -i\Delta_{+-}(x,x')\right]\nonumber\\&&+ \frac{i\alpha \beta}{2} \frac{H^{2-\e}}{2^{3-\e} \pi^{2-\e/2}} \frac{\Gamma(3-\e)}{\Gamma(2-\e/2)} \int d^d x' a'^d \left(\frac{1}{\e} +\ln a'\right) \left[ i\Delta_{++}(x,x') -i\Delta_{+-}(x,x')\right] \nonumber\\&&+ i\alpha^2 \int d^d x' a'^d \left[ i\Delta_{++}(x,x') -i\Delta_{+-}(x,x')\right]
\label{l49}
\end{eqnarray} 
where $\alpha$ is the one loop tadpole counterterm,  \ref{b2}. Substituting this into the above equation, we find the divergences cancel to give,
\begin{eqnarray}
\frac{i\beta^2 H^{4-2\e}}{2^{8-2\e} \pi^{4-\e}} \frac{\Gamma^2(3-\e)}{\Gamma^2(2-\e/2)} \ln a \int d^d x' a'^d\, \ln a' \, \left[ i\Delta_{++}(x,x') -i\Delta_{+-}(x,x')\right]
\label{l50}
\end{eqnarray} 
Using \ref{l7}, \ref{l8}, it is easy to check that the above integral is ultraviolet finite. The detail of computing it  is given at the first part of \ref{C}.  Using \ref{c16}, the above integral is  evaluated to be
\begin{eqnarray}
\frac{\beta^2 H^2}{2^6 \pi^4} \frac{\ln a}{3} \left[ \frac19 - \frac{1}{9a^3} - \frac13 \ln a +\frac12 \ln^2 a \right]+{\cal O}(\e)
\label{l52}
\end{eqnarray} 
\vskip .5cm

Let us now evaluate the second integral of \ref{l48}, along with the contribution from the one loop mass renormalisation counterterm (second row of \ref{f3}) given by \ref{b5}. It equals 
\begin{eqnarray}
\frac{i\beta^2}{6} \int d^d x' a'^d\,\left[ (i \Delta_{++}(x,x'))^3 -(i\Delta_{+-}(x,x'))^3\right]  - \, \frac{\beta^2 \mu^{-\e} \Gamma(1-\e/2)}{2^5 \pi^{2-\e/2} \e (1-\e)}\, i\Delta (x,x) 
\label{l53}
\end{eqnarray} 
Note that such cube of the propagators  was computed earlier in~\cite{Brunier:2004sb} in the context  of $\phi^4$ theory. Here in addition we need to perform the volume integral, detailed in~\ref{C}. \ref{l53} is then evaluated to be,
 \begin{eqnarray}
&& - \frac{\beta^2 H^2 \mu^{-2\e} \Gamma^2(1-\e/2) (1+\e/2) (1+\e/4)}{3\times 2^9 \pi^{4-\e} (1-3\e/2)(1-\e)(1-3\e/4) \e} +\frac{\beta^2 \mu^{-\e}H^{2-\e}\Gamma(1-\e)}{2^{7-\e} \pi^{4-\e} \e^2} \left(1- \left(\frac{2\mu}{H} \right)^{-\e}\frac{(1-\e/4)\Gamma(2-\e/2)}{(1-3\e/2)}\right) \nonumber\\
&& - \frac{\beta^2 H^2 }{2^7 \pi^4}  \left(2 \zeta(3)+\frac{799}{864}+\frac{11\pi^2}{18}+2 \ln \frac{2\mu e^{-3/4}}{H} \ln \frac{2\mu }{e H} -\frac52 \ln \frac{2\mu}{\sqrt{e}H} + \ln^2 \frac{2\mu}{\sqrt{e}H} \right. \nonumber\\ && \left.+4 \sum \frac{(n^2-2n-1+2/n)}{n(n-1)^2(n-2)^2(n+1)^2} -\sum\frac{2 }{3mn(m+n) (m+n+1) (m+n+2) (m+n+3)}\right) \nonumber\\
&&+\frac{\beta^2H^{2}}{2^{6}\pi^{4}}\,\ln a \left\{\left(\frac{2\pi}{\mu H}\right)^{\e}\frac{1}{\e}  \left(\frac{3\Gamma(1-\e)}{2}  -  \left(\frac{2\mu}{H} \right)^{-\e}\frac{\Gamma(1-\e/2)\Gamma(2-\e/2)(1-\e/4)}{(1-3\e/2)}\right) + \ln \frac{2\mu}{H}\right\}\nonumber\\
\nonumber\\ && + \frac{\beta^2 H^2 }{2^7 \pi^4} \left[\left(\frac{\pi^2}{6}- \frac{977}{144} \right)\frac{1}{a} - \left(\frac{\pi^2}{12}+\frac{835}{288} \right)\frac{1}{a^2} + \left(\frac{241}{432}-\frac{\pi^2}{18} \right) \frac{1}{a^3}  + \left(\frac{677}{1296}+\frac{\pi^2}{54} \right)\ln a+ \frac{71}{36}\ln^2a +\frac{1}{27}\ln^3 a \right. \nonumber\\ && \left. + \sum\left\{\left(\frac{198 n^4+387 n^3-171 n^2-414 n-38}{108 n^2}- \ln \frac{2\mu}{H} \frac{(n^2-1)(n+2)}{n}\right) a^{-n} \right.\right. \nonumber\\ &&\left.\left.  + \left(\frac{(n+2)(n+3)}{3}\ln \frac{2\mu}{H}-\frac{11 n^5+77 n^4+187 n^3+151 n^2-25 n-163}{18 n (n+1)^2} \right)a^{-n-1}  -\frac{19 n^2-27 n-274}{18 n (n+1) (n+2)^2} a^{-n-2} \right.\right. \nonumber\\ && \left.\left. 
+ \frac{19 a^{-n-3}}{54 n(n+3)} -\left(\frac{11}{6}- \ln \frac{2\mu}{H}\right)\left( (n+1)(n-2)a^{-n+1}-\frac{(n-1)(n-3)a^{-n+2}}{3}\right)  \right\} + \sum \frac{1}{mn} \left\{ \frac{a^{-m-n-3}}{9(m+n+3)}  \right. \right.\nonumber\\ && \left.\left. +\frac{11 a^{-m-n-2}}{3(m+n+2)} - \frac{(m+n)(m+n+1)(m+n-2)}{2}a^{-m-n+1} +\frac{(m+n)(m+n-1)(m+n-3)}{6}a^{-m-n+2}   \right. \right. \nonumber\\ && \left.\left.
-\left(\frac{(m+n) (m+n+2) (m+n+3)}{2}+\frac{11}{(m+n+1)}\right)\frac{a^{-m-n-1}}{3} \right. \right. \nonumber\\ && \left.\left.
+ \left( \frac{(m+n-1) (m+n+1) (m+n+2)}{2}-\frac{1}{9 (m+n)}\right)a^{-m-n}
\right\} \right] \qquad 
\label{c10''}
\end{eqnarray}
 Combining now the above with \ref{l52}, we obtain the unrenormalised two loop expression,
%
\begin{eqnarray}
&& - \frac{\beta^2 H^2 \mu^{-2\e} \Gamma^2(1-\e/2) (1+\e/2) (1+\e/4)}{3\times 2^9 \pi^{4-\e} (1-3\e/2)(1-\e)(1-3\e/4) \e} +\frac{\beta^2 \mu^{-\e}H^{2-\e}\Gamma(1-\e)}{2^{7-\e} \pi^{4-\e} \e^2} \left(1- \left(\frac{2\mu}{H} \right)^{-\e}\frac{(1-\e/4)\Gamma(2-\e/2)}{(1-3\e/2)}\right) \nonumber\\
&& - \frac{\beta^2 H^2 }{2^7 \pi^4}  \left(2 \zeta(3)-\frac{1043}{864}+\frac{17\pi^2}{18}+2 \ln \frac{2\mu e^{-3/4}}{H} \ln \frac{2\mu }{e H} -\frac52 \ln \frac{2\mu}{\sqrt{e}H} + \ln^2 \frac{2\mu}{\sqrt{e}H} \right. \nonumber\\ && \left. -\sum_{n,m=1}^{\infty}\frac{2 }{3mn(m+n) (m+n+1) (m+n+2) (m+n+3)}\right) \nonumber\\
&&+\frac{\beta^2H^{2}}{2^{6}\pi^{4}}\,\ln a \left\{\left(\frac{2\pi}{\mu H}\right)^{\e}\frac{1}{\e}  \left(\frac{3\Gamma(1-\e)}{2}  -  \left(\frac{2\mu}{H} \right)^{-\e}\frac{\Gamma(1-\e/2)\Gamma(2-\e/2)(1-\e/4)}{(1-3\e/2)}\right) + \ln \frac{2\mu}{H}\right\}\nonumber\\
\nonumber\\ && + \frac{\beta^2 H^2 }{2^7 \pi^4} \left[\left(\frac{\pi^2}{6}- \frac{977}{144} \right)\frac{1}{a} - \left(\frac{\pi^2}{12}+\frac{475}{288} \right)\frac{1}{a^2} + \left(\frac{121}{432}-\frac{\pi^2}{18} \right) \frac{1}{a^3}  + \left(\frac{2933}{1296}+\frac{\pi^2}{54}  \right)\ln a+ \frac{7}{4}\ln^2a +\frac{10}{27}\ln^3 a \right. \nonumber\\ && \left. -\frac{2\ln a}{27 a^3}+ \sum_{n=1}^{\infty}\left\{\left(\frac{198 n^4+387 n^3-171 n^2-414 n-38}{108 n^2}-  \frac{(n^2-1)(n+2)}{n}\ln \frac{2\mu}{H}\right) a^{-n} \right.\right. \nonumber\\ &&\left.\left.  + \left(\frac{(n+2)(n+3)}{3}\ln \frac{2\mu}{H}-\frac{11 n^5+77 n^4+187 n^3+151 n^2-25 n-163}{18 n (n+1)^2} \right)a^{-n-1}  -\frac{19 n^2-27 n-274}{18 n (n+1) (n+2)^2} a^{-n-2} \right.\right. \nonumber\\ && \left.\left. 
+ \frac{19 a^{-n-3}}{54 n(n+3)} -\left(\frac{11}{6}- \ln \frac{2\mu}{H}\right)\left( (n+1)(n-2)a^{-n+1}-\frac{(n-1)(n-3)a^{-n+2}}{3}\right)  \right\} + \sum_{n,m=1}^{\infty} \frac{1}{mn} \left\{ \frac{a^{-m-n-3}}{9(m+n+3)}  \right. \right.\nonumber\\ && \left.\left. +\frac{11 a^{-m-n-2}}{3(m+n+2)} - \frac{(m+n)(m+n+1)(m+n-2)}{2}a^{-m-n+1} +\frac{(m+n)(m+n-1)(m+n-3)}{6}a^{-m-n+2}   \right. \right. \nonumber\\ && \left.\left.
-\left(\frac{(m+n) (m+n+2) (m+n+3)}{2}+\frac{11}{(m+n+1)}\right)\frac{a^{-m-n-1}}{3}  \right. \right. \nonumber\\ && \left.\left.+ \left( \frac{(m+n-1) (m+n+1) (m+n+2)}{2}-\frac{1}{9 (m+n)}\right)a^{-m-n}
\right\} \right]
\label{c10'}
\end{eqnarray}
Thus the above expression, when multiplied with $g_{\mu\nu}$ yields the two loop vacuum contribution of the cubic potential to the energy-momentum tensor, \ref{l48}.

None of the  summations appearing above need any regularisation. 
Some of them  can be exactly evaluated using, e.g. Mathematica. However, all of them are either constant or contain negative powers of the scale factor, thereby the latter drop out at late cosmological times, and hence we shall not bother ourselves in evaluating them.  In particular, we also note that many of the summations diverge at the initial time $t=0$ ($a=1$). For example, the first summation appearing in the sixth row of \ref{c10'} is evaluated to be
\begin{eqnarray}
&&\sum_{n=1}^{\infty}\frac{19 n^4+38 n^3-17 n^2-41 n-38}{108 n^2 a^n}\nonumber\\&&=
\frac{\left(38(1- a^3)+114 a(a-1) \right)\text{Li}_2\frac{1}{a}-41(1- a^3) \ln \frac{a-1}{a}-123 a(a-1) \ln \frac{a-1}{a}-17+15 a+40 a^2}{108 (a-1)^3}\qquad
\label{sum}
\end{eqnarray}
where Li is the polylog function~\cite{GR}.  The above expression is bounded everywhere except as $a\to 1$, and vanishes at late times, $a\gg 1$. Note that the summations corresponding to the quartic theory appearing in \ref{l47} also show similar feature. This can be understood as follows~\cite{Kahya:2009sz}. In the flat spacetime, the initial states are defined in the asymptotic past and they are essentially free states. The perturbation is `turned on and off' in between the asymptotic past and future, and one then knows the complete evolution of the states. In the cosmological scenario however, we define our initial states at some finite initial time (i.e., $t\to 0$, or $a\to 1$ in our case), and evolve it up to some finite final time only. Also in particular, we do not have the freedom to turn on and off the interaction here, as of the flat spacetime. Such interactions create loops where virtual particles are created and hence may amplify the overall particle creation in the expanding background.  These reasons should be  attributed to the divergence of \ref{sum}, showing the blueshift of the quantum corrections on the initial Cauchy surface.  Such divergences seem to be tackled by perturbatively correcting the initial Bunch-Davies vacuum state, as was pointed out in~\cite{Kahya:2009sz}. However, since none of such terms yield any significant secular contributions at late times we are interested in,  we may safely ignore them for our current purpose.

In order to renormalise \ref{c10'}, we consider the  part that gives the leading late time contribution (i.e., the part without derivatives, which yield non-dynamical terms) from the conformal counterterm, \ref{l42},
$$ \langle T_{\mu\nu}^{\delta \xi}\rangle= -\frac{\delta \xi H^{4-\e}}{2^{2-\e} \pi^{2-\e/2}}\frac{\Gamma(4-\e)}{\Gamma(2-\e/2)}\left(\frac{1}{\e} +\ln a\right) g_{\mu\nu} $$
Choosing
$$\delta \xi =  \frac{\beta^2 H^{-2+\e}}{2^{4+\e} \pi^{2+\e/2}}\frac{\Gamma(2-\e/2)}{\Gamma(4-\e)}\left[\left(\frac{2\pi}{\mu H}\right)^{\e}\frac{1}{\e}  \left(\frac{3\Gamma(1-\e)}{2}  -  \left(\frac{2\mu}{H} \right)^{-\e}\frac{\Gamma(1-\e/2)\Gamma(2-\e/2)(1-\e/4)}{(1-3\e/2)}\right) + \ln \frac{2\mu}{H}\right]$$
removes the fourth row of \ref{c10'}. Note from \ref{l42} that we also generate a ${\cal O}(\e^{-1})$ contribution containing $\delta_{\mu}^0\delta_{\nu}^0$ had we considered the derivatives as well. However such tensor structure indicates that they will be necessary to renormalise the kinetic term, which we have not considered here. We next  choose the cosmological counterterm
\begin{eqnarray*}
&&\frac{\delta \Lambda}{8\pi G}=- \frac{\beta^2 H^2 \mu^{-2\e} \Gamma^2(1-\e/2) (1+\e/2) (1+\e/4)}{3\times 2^9 \pi^{4-\e} (1-3\e/2)(1-\e)(1-3\e/4) \e} +\frac{\beta^2 \mu^{-\e}H^{2-\e}\Gamma(1-\e)}{2^{7-\e} \pi^{4-\e} \e^2} \left(1- \left(\frac{2\mu}{H} \right)^{-\e}\frac{(1-\e/4)\Gamma(2-\e/2)}{(1-3\e/2)}\right) \nonumber\\
&& - \frac{\beta^2 H^2 }{2^7 \pi^4}  \left(2 \zeta(3)-\frac{1043}{864}+\frac{17\pi^2}{18}+2 \ln \frac{2\mu e^{-3/4}}{H} \ln \frac{2\mu }{e H} -\frac52 \ln \frac{2\mu}{\sqrt{e}H} + \ln^2 \frac{2\mu}{\sqrt{e}H}\right. \nonumber\\ && \left.-\sum_{n,m=1}\frac{2 }{3mn(m+n) (m+n+1) (m+n+2) (m+n+3)}\right)\nonumber\\&&-\frac{\beta^2H^2}{2^6\pi^4\e}\left[\left(\frac{2\pi}{\mu H}\right)^{\e}\frac{1}{\e}  \left(\frac{3\Gamma(1-\e)}{2}  -  \left(\frac{2\mu}{H} \right)^{-\e}\frac{\Gamma(1-\e/2)\Gamma(2-\e/2)(1-\e/4)}{(1-3\e/2)}\right) + \ln \frac{2\mu}{H}\right],
\end{eqnarray*}
removing the first three lines of \ref{c10'} as well as the ${\cal O}(\e^{-2})$ divergence generated by $\delta \xi$. Putting these all in together, we read off from \ref{c10'} the leading late time two loop secular contribution for the cubic sector,
\begin{eqnarray}
\langle T_{\mu\nu}\rangle_{\beta,~{\rm Ren.}}\big\vert_{a\gg 1} \approx  \frac{\beta^2 H^2}{2^6 \pi^4}\frac{5\ln^3 a}{27} g_{\mu\nu}
\label{finalcu}
\end{eqnarray}
This behaviour is more vicious compared to that of the quartic sector,~\ref{l47}. This is expected, as the latter corresponds to lesser number of internal lines in the vacuum diagrams. We note also that \ref{finalcu} corresponds to negative energy density and positive pressure, in contrast  to  the quartic sector~\ref{l47}. There is no reason to believe that even one computes some higher order corrections and attempts some resummation, we would obtain any bounded result,     
corresponding to the fact that the $\phi^3$ potential is unbounded from below. Thus we also need to  add to it a positive, quartic potential  in order do any sensible physics. We could also have added \ref{finalcu} with \ref{l47} and have tried some resummation to compute a non-perturbative $\langle V(\phi)\rangle$. However,  intuitively it seems that in order to obtain a better result we should go to further higher orders where $\beta$ and $\lambda$ are co-existing.  We reserve this problem for a future work. Instead, we shall compute below  $\langle \phi \rangle$ in such quartic-cubic potential, in order to find out a  non-perturbative, finite expression at late times, effectively behaving as a cosmological constant.

\section{$\langle \phi\rangle$ in $\lambda \phi^4/4!+\beta \phi^3/3!$ } \label{tadpole}

We shall compute below  $\langle \phi(x) \rangle$ perturbatively up to two loop order, ${\cal O}(\beta)$ and ${\cal O}(\lambda\beta)$. Using these, we shall find a resummed expression afterward.  Computation of $\langle \phi(x) \rangle$ is essentially related to computing the tadpoles. The one-loop contribution comes from the cubic self interaction plus the one loop tadpole counterterm, given by the first row of \ref{f4} with the external point at $x$ attached,
\begin{eqnarray}
\langle\phi\rangle_{1-{\rm loop,}\,\phi^3}= -\frac{i\beta}{2}\int d^d x' a'^d\left(i\Delta_{++}(x,x')-i\Delta_{+-}(x,x') \right)i\Delta(x',x') -i\alpha\int d^d x' a'^d\left(i\Delta_{++}(x,x')-i\Delta_{+-}(x,x') \right)
\label{l65}
\end{eqnarray}
The divergence in $i\Delta(x',x')$, \ref{l13}, gets cancelled by the counterterm $\alpha$, \ref{b2}. The remaining integral is ultraviolet finite and has been evaluated in \ref{C} (\ref{c10}, \ref{c16}). We have 
\begin{eqnarray}
\langle\phi\rangle_{1-{\rm loop,}\,\phi^3}\big\vert_{\rm Ren.}= -\frac{\beta}{2^3 \times 3\pi^2}\left[\frac{1}{9}-\frac{1}{9a^3}-\frac{\ln a}{3}+\frac{\ln^2 a}{2} \right]
\label{l66}
\end{eqnarray}
Note that the flat space limit (equivalently, the initial time, $t=0$) corresponds to $a=1$, in which case the above expectation value vanishes. 
	\begin{figure}[h!]
		\includegraphics[height=1.5cm]{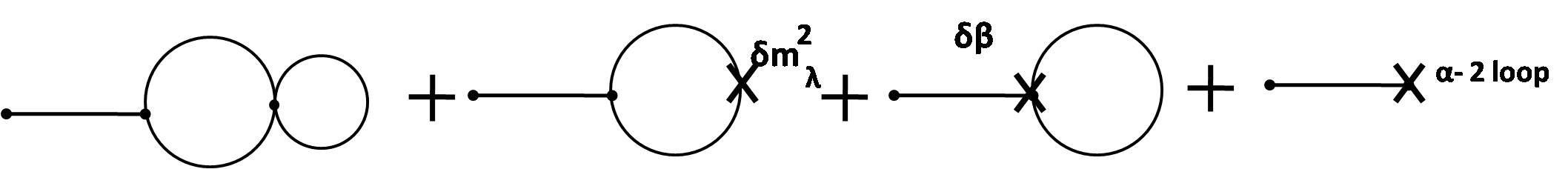}\centering \hskip 2cm
		\includegraphics[height=1.5cm]{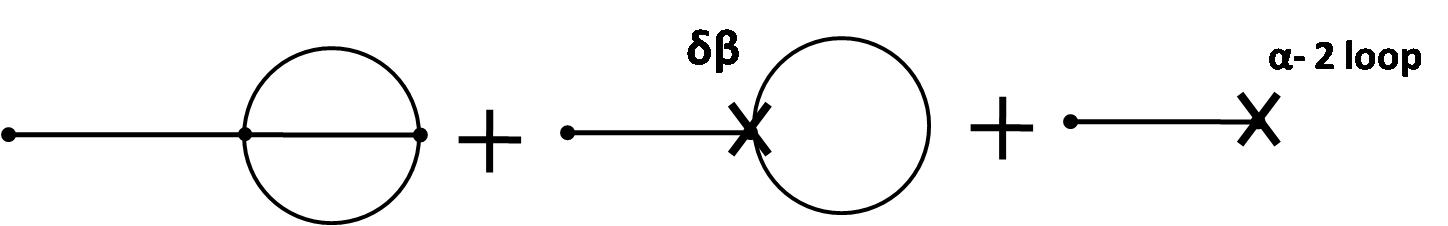}\centering \hskip 2cm
		\includegraphics[height=1.5cm]{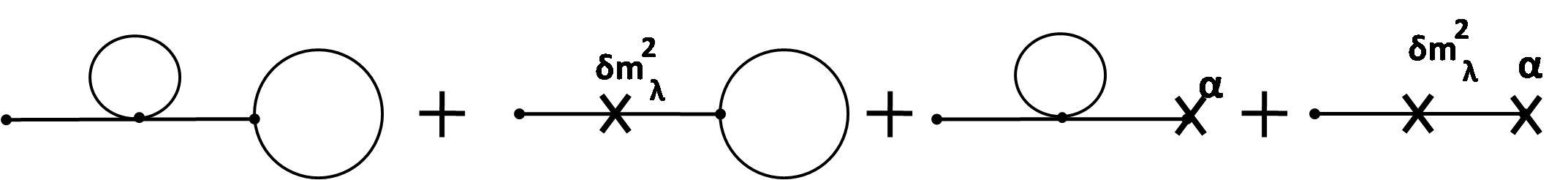}
		\caption{ \footnotesize Tadpole diagrams corresponding respectively (row-wise) to \ref{l67}, \ref{l73} and \ref{l79a1}.}
		\label{f5}
	\end{figure}
Let us now compute $\langle \phi(x) \rangle$ at the next to the leading order, in which case we have contributions from both quartic and cubic self interaction. There are three such processes, the first  is given by the first row of \ref{f5}, 
\begin{eqnarray}
\langle\phi\rangle_{2-{\rm loop,}\,\phi^3+\phi^4}^{(1)}= -\frac{\beta \lambda}{4}\int d^d x' d^d x'' (a' a'')^d i\Delta (x'',x'') \left[i\Delta_{++}(x,x')\left(i\Delta^2_{++}(x',x'')-i\Delta^2_{+-}(x',x'') \right) \right. \nonumber\\  \left.+ i\Delta_{+-}(x,x')\left(i\Delta^2_{--}(x',x'')-i\Delta^2_{-+}(x',x'') \right)  \right]\nonumber\\
=- \frac{\beta \lambda H^{2-\e}}{2^{5-\e}\pi^{2-\e/2}}\frac{\Gamma(3-\e)}{\Gamma(2-\e/2)}\int d^d x' d^d x'' (a' a'')^d \left(\frac{1}{\e}+\ln a'' \right)\left[i\Delta_{++}(x,x')\left(i\Delta^2_{++}(x',x'')-i\Delta^2_{+-}(x',x'') \right) \right. \nonumber\\  \left.+ i\Delta_{+-}(x,x')\left(i\Delta^2_{--}(x',x'')-i\Delta^2_{-+}(x',x'') \right)  \right]
\label{l67}
\end{eqnarray}
where we have used  \ref{l13} in the second equality. Note from \ref{feyn} that the the above integral vanishes for $\eta' \lesssim \eta''$. We also note that $\left(i\Delta^2_{++}(x',x'')-i\Delta^2_{+-}(x',x'') \right)=\left(i\Delta^2_{--}(x',x'')-i\Delta^2_{-+}(x',x'') \right)^{\star}$. This implies the temporal hierarchy,
$$\eta \gtrsim \eta' \gtrsim \eta''$$
in \ref{l67}. Let us first deal with the divergent part of it. We now add to the above the process involving mass counterterm for the $\phi^4$ theory, \ref{l47}, and the cubic interaction, 
\begin{eqnarray}
-\frac{\beta \delta m^2_{\lambda}}{2} \int d^d x' d^d x'' (a'a'')^d \left[i\Delta_{++}(x,x')\left(i\Delta^2_{++}(x',x'')-i\Delta^2_{+-}(x',x'') \right) \right. \nonumber\\  \left.+ i\Delta_{+-}(x,x')\left(i\Delta^2_{--}(x',x'')-i\Delta^2_{-+}(x',x'') \right)  \right]
\label{l68'}
\end{eqnarray}
\ref{l47'} shows that the above cancels the divergence of \ref{l67}. However, further divergence comes from the square of the propagator, attached to a $\delta$-function,  via \ref{b3}, \ref{l23} and \ref{l29}. Using them and recalling $d=4-\e$, we have for the divergent part of \ref{l67}
\begin{eqnarray}
&&\langle\phi\rangle^{(1)}
_{2-{\rm loop,}\,\phi^3+\phi^4}\vert_{\rm div.}=\frac{i\mu^{-\e}\beta \lambda H^{2-\e}\Gamma(3-\e)}{2^{8-\e}\pi^{4-\e}\e(1-\e)(1-\e/2)}\int d^d x' a'^{(4-\e)+\e} \ln a' \left(i\Delta_{++}(x,x')-i\Delta_{+-}(x,x') \right)\nonumber\\
&&=\frac{i\mu^{-\e}\beta \lambda H^{2-\e}\Gamma(3-\e)}{2^{8-\e}\pi^{4-\e}\e(1-\e)(1-\e/2)}\int d^d x' a'^{d}  \ln a' \left(i\Delta_{++}(x,x')-i\Delta_{+-}(x,x') \right)\nonumber\\&&+ \frac{i\beta \lambda H^2}{2^7\pi^4}\int d^4 x' a'^4 \ln^2 a' \left(i\Delta_{++}(x,x')-i\Delta_{+-}(x,x') \right)
\label{l67'}
\end{eqnarray}
The above divergence cannot be cancelled by the mass counterterm of the cubic theory, as it is ${\cal O}(\beta^2)$, \ref{b5}, whereas we are presently working at ${\cal O}(\lambda\beta)$. Note also that such divergence is not present in flat spacetime as $\ln a$ must be set to zero in that case. Thus in the flat spacetime, \ref{l68'} would have been suffice to deal with the divergence of \ref{l67}, eventually making the latter entirely vanishing. Hence we further introduce  the contribution from the cubic vertex counterterm,
\begin{eqnarray}
-\frac{i\delta \beta H^{2-\e} \Gamma(3-\e)}{2^{4-\e}\pi^{2-\e/2} \Gamma(2-\e/2)}\int d^d x' a'^d\left(i\Delta_{++}(x,x')-i\Delta_{+-}(x,x') \right)\left(\frac{1}{\e}+\ln a' \right)
\label{l68}
\end{eqnarray}
where we have used \ref{l13}. Comparison of \ref{l67'} and \ref{l68} shows that we must choose
\begin{eqnarray}
\delta \beta = \frac{\mu^{-\e}\beta \lambda \Gamma(1-\e/2)}{2^4 \pi^{2-\e/2}\e(1-\e)}
\label{l69}
\end{eqnarray}
to get rid of the divergence in the second line of \ref{l67'}. 
$\delta \beta$ however, introduces an ${\cal O}(\e^{-2})$ divergence which can be canceled  by the two loop tadpole counterterm,
\begin{eqnarray}
\alpha_{\rm 2-loop} =- \frac{\mu^{-\e}\beta \lambda H^{2-\e}\Gamma(1-\e)}{2^{7-\e}\pi^{4-\e}\e^2}
\label{l70}
\end{eqnarray}

The rest of the integral \ref{l67} is ultraviolet finite. We next compute the finite part of \ref{l67'} (the last line), using the results of \ref{A} and at the beginning of \ref{C}, given by
\begin{eqnarray}
\frac{\beta \lambda}{2^7\times 3\pi^4}\left[\frac{2}{27 a^3}+\frac{\ln^3a}{3}-\frac{\ln^2a}{3}+\frac{2 \ln a}{9}-\frac{2}{27}\right]
\label{l71}
\end{eqnarray}

The rest of the integral \ref{l67} (the last line) can be evaluated using the methods of \ref{A}, \ref{b3}, \ref{c11}. The actual expression is very lengthy and we shall quote only the secular logarithm terms,
\begin{eqnarray}
\frac{\beta \lambda}{2^7\times 3\pi^4}\left[\frac{8}{27}\ln^2 a -\frac{8}{27}\ln^3 a + \frac{\ln^4 a}{9} \right]
\label{l72}
\end{eqnarray}
Combining the above with the secular terms of \ref{l71}, we have the renormalised and leading late time expression for \ref{l67}
\be
\langle\phi\rangle_{2-{\rm loop,}\,\phi^3+\phi^4}^{(1)}\big\vert_{a\gg 1, {\rm Ren.}}=\frac{\beta \lambda}{2^7\times 3\pi^4} \left[\frac{\ln^4 a}{9}+\frac{\ln^3 a}{27}  +{\cal O}(\ln^2a )\right]
\label{one}
\ee
\bigskip

\noindent
Let us now compute the second diagram, the second row of \ref{f5}. It equals 
\begin{eqnarray}
\langle\phi\rangle_{2-{\rm loop,}\,\phi^3+\phi^4}^{(2)}=&&-\frac{\beta \lambda}{6} \int d^d x' d^d x'' (a'a'')^d \left[i\Delta_{++}(x,x')(i\Delta_{++}^3(x',x'') -i\Delta_{+-}^3(x',x'')) \right. \nonumber\\ && \left.+ i\Delta_{+-}(x,x')(i\Delta_{--}^3(x',x'') -i\Delta_{-+}^3(x',x'')) \right]
\label{l73}
\end{eqnarray}
Using \ref{c3} (with the replacement, $\beta^2 \to i\beta \lambda$) and the fact that $(i\Delta_{++}^3(x',x'') -i\Delta_{+-}^3(x',x''))= (i\Delta_{--}^3(x',x'') -i\Delta_{-+}^3(x',x''))^{\star}$, we evaluate the above integral upto the leading secular order ${\cal O}(\ln^3 a)$ and ${\cal O}(\ln^4 a)$. As of \ref{l67}, we also have in this case the temporal hierarchy, $\eta\gtrsim \eta' \gtrsim \eta''$. Let us  first consider the contribution from the first two terms on the right hand side of \ref{c3}. This part will give divergence and hence requires renormalisation.  We have
\begin{eqnarray}
&&-\frac{i\beta \lambda H^2 \mu^{-2\e}}{ 2^{9} \times3\pi^{4-\e}} \frac{\Gamma^2(1-\e/2)(1+\e/2)(1+\e/4)}{(1-3\e/2)(1-\e)(1-3\e/4)\e} \int d^d x' a'^d(i\Delta_{++}(x,x')-i\Delta_{+-}(x,x')) \nonumber\\
&&-\frac{i\beta \lambda \mu^{-\e} H^{2-\e}}{2^{7-\e}\pi^{4-\e}}\int d^d x' a'^d \left[ \frac{1}{\e^2}\left(\frac{(1-\e/2)(1-\e/4)\Gamma^2(1-\e/2)}{(1-3\e/2)}\left(\frac{2\mu}{H}\right)^{-\e}-2\Gamma(1-\e) \right)\right.\nonumber\\ &&\left.+\frac{2}{\e}\left( \frac{(1-\e/2)(1-\e/4)\Gamma^2(1-\e/2)}{(1-3\e/2)}\left(\frac{2\mu}{H}\right)^{-\e}-2\Gamma(1-\e)\right)\ln a'\right](i\Delta_{++}(x,x')-i\Delta_{+-}(x,x')) \nonumber\\
&&+ \frac{3i\beta \lambda H^2}{2^8 \pi^4} \int d^4 x' a'^4 \ln^2a' \left(i\Delta_{++}(x,x')-i\Delta_{+-}(x,x') \right)
\label{l74}
\end{eqnarray}
The integral in the last line is ultraviolet finite which contributes to the secular growth at late times. In order to get rid of the ${\cal O}(\e^{-1})$ divergence of the above expression, we add as earlier the contribution from the cubic vertex counterterm, \ref{l68}, with the choice
\begin{eqnarray}
\delta \beta = -\frac{\mu^{-\e} \beta \lambda}{2^2 \pi^{2-\e/2} \e}\frac{\Gamma(2-\e/2)}{\Gamma(3-\e)} \left( \frac{(1-\e/2)(1-\e/4)\Gamma^2(1-\e/2)}{(1-3\e/2)}\left(\frac{2\mu}{H}\right)^{-\e}-2\Gamma(1-\e)\right)
\label{l75}
\end{eqnarray}
The remaining ${\cal O}(\e^{-2})$ and ${\cal O}(\e^{-1})$ divergences are canceled by the choice of the  two loop tadpole counterterm,
\begin{eqnarray}
\alpha_{\rm 2-loop} = &&-\frac{\mu^{-\e} \beta \lambda  H^{2-\e}}{2^{6-\e} \pi^{6-\e}\e^2} \left( \frac{(1-\e/2)(1-\e/4)\Gamma^2(1-\e/2)}{(1-3\e/2)}\left(\frac{2\mu}{H}\right)^{-\e}-2\Gamma(1-\e)\right)\nonumber\\&&+\frac{\beta \lambda H^2 \mu^{-2\e}}{2^{9} \times 3\pi^{4-\e} \e}\frac{\Gamma^2(1-\e/2)(1+\e/2)(1+\e/4)}{(1-3\e/2)(1-\e)(1-3\e/4)}
\label{l76}
\end{eqnarray}
The last integral of \ref{l74} can be evaluated in a similar manner as of \ref{C},  yielding its renormalised  leading secular contribution,
\begin{eqnarray}
\frac{\beta \lambda }{2^8 \times 3 \pi^4} \left[\ln^3 a +{\cal O}(\ln^2a)\right]
\label{l77}
\end{eqnarray}

The remaining integrals of \ref{l73} are ultraviolet finite, and can be evaluated using \ref{c5}, \ref{c7}, \ref{c8} and \ref{c9} of \ref{C}. The   contribution from the $x''$-integral which becomes leading order at late times is given by,
\begin{eqnarray}
\frac{i\beta \lambda H^2}{2^7\pi^4}\left[ \frac{35 \ln^2 a'}{36}+\frac{\ln^3 a'}{27}\right]
\label{l78}
\end{eqnarray}
We substitute this back into \ref{l73} and use \ref{c11} to perform the remaining $x'$-integral.  The leading secular contribution is found to be,
\begin{eqnarray}
\frac{\beta \lambda}{2^7\pi^4}\left[ \frac{\ln^4 a}{324} +\frac{101\ln^3a}{972}+{\cal O}(\ln^2 a)\right]
\label{l79}
\end{eqnarray}
Combining the above with \ref{l77}, we find the renormalised, leading late time secular growth corresponding to \ref{l73}
\begin{eqnarray}
\langle\phi\rangle_{2-{\rm loop,}\,\phi^3+\phi^4}^{(2)}\big\vert_{{\rm Ren.}, a\gg1}=\frac{\beta \lambda}{2^7\times 3\pi^4}\left[ \frac{\ln^4 a}{108} +\frac{263\ln^3a}{324}+{\cal O}(\ln^2 a)\right]
\label{l79'}
\end{eqnarray}

The third process is given by the third row of \ref{f5},
\begin{eqnarray}
\langle\phi\rangle_{2-{\rm loop,}\,\phi^3+\phi^4}^{(3)}=&&
-\frac{\beta \lambda }{4} \int (a'a'')^d d^d x' d^d x'' i\Delta(x',x')i\Delta(x'',x'') \left[i\Delta_{++}(x,x')(i\Delta_{++}(x',x'') - i\Delta_{+-}(x',x'')) \right. \nonumber\\ && \left.+ i\Delta_{+-}(x,x')(i\Delta_{--}(x',x'') - i\Delta_{-+}(x',x'')) \right]
\label{l79a1}
\end{eqnarray}
whereas the contributions from the one-loop tadpole and the one-loop mass renormalisation counterterms to be added with the above reads
\begin{eqnarray}
&&-\frac{\alpha \lambda}{2} \int (a'a'')^d d^d x' d^d x'' i\Delta(x',x')\left[i\Delta_{++}(x,x')(i\Delta_{++}(x',x'') - i\Delta_{+-}(x',x'')) \right. \nonumber\\ && \left.+ i\Delta_{+-}(x,x')(i\Delta_{--}(x',x'') - i\Delta_{-+}(x',x'')) \right] \nonumber\\
&&-\frac{\beta \delta m^2_{\lambda}}{2} \int (a'a'')^d d^d x' d^d x'' i\Delta(x'',x'')\left[i\Delta_{++}(x,x')(i\Delta_{++}(x',x'')- i\Delta_{-+}(x',x'')) \right. \nonumber\\ && \left.+ i\Delta_{+-}(x,x')(i\Delta_{--}(x',x'') - i\Delta_{-+}(x',x'')) \right] \nonumber\\
&&-\alpha\delta m^2_{\lambda} \int (a'a'')^d d^d x' d^d x'' \left[i\Delta_{++}(x,x')(i\Delta_{++}(x',x'') - i\Delta_{+-}(x',x'')) \right. \nonumber\\ && \left.+ i\Delta_{+-}(x,x')(i\Delta_{--}(x',x'') - i\Delta_{-+}(x',x'')) \right] 
\label{l79a2}
\end{eqnarray}
where $\delta m^2_{\lambda} $ and $\alpha$ are given by \ref{l47'}, \ref{b2}. Using then \ref{l13} it is easy to see that the above cancels all the divergences of \ref{l79a1}, yielding the ultraviolet finite integral
\begin{eqnarray}
\langle\phi\rangle_{2-{\rm loop,}\,\phi^3+\phi^4}^{(3)}=&&
-\frac{\beta \lambda H^4 }{2^6 \pi^4} \int (a'a'')^4 d^4 x' d^4 x'' \ln a' \ln a'' \left[i\Delta_{++}(x,x')(i\Delta_{++}(x',x'') - i\Delta_{+-}(x',x'')) \right. \nonumber\\ && \left.+ i\Delta_{+-}(x,x')(i\Delta_{--}(x',x'') - i\Delta_{-+}(x',x'')) \right]
\label{l79a3}
\end{eqnarray}
Following the method described at the beginning of \ref{C}, the above integral is evaluated as
\begin{eqnarray}
\langle\phi\rangle_{2-{\rm loop,}\,\phi^3+\phi^4}^{(3)}\big\vert_{{\rm Ren.}, a\gg1}=
\frac{\beta \lambda  }{2^7\times  3\pi^4} \left[ \frac{\ln^4a}{12}-\frac{37}{54}\ln^3 a + {\cal O}(\ln^2a) \right]
\label{l79a4}
\end{eqnarray}

Adding now \ref{one}, \ref{l79'} and \ref{l79a4},
 we finally obtained the two loop ${\cal O}(\lambda \beta) $ renormalised and leading late time result for $\langle\phi\rangle$, 
\begin{eqnarray}
\langle\phi\rangle_{2-{\rm loop,}\,\phi^3+\phi^4}\big\vert_{{\rm Ren.}, a\gg1} = \frac{\beta \lambda}{2^7 \times 3\pi^4}\left[\frac{11}{54}\ln^4 a+ \frac{53 }{324}\ln^3 a +{\cal O}(\ln^2a) \right],
\label{l80}
\end{eqnarray}
 indicating the breakdown of the perturbation theory after sufficient $e$-foldings, and necessitating some kind of resummation.

\subsection{Resummation}\label{resum}

The traditional dynamical renormalisation group method for resummation may not always work in the inflationary scenario. This is because in this case even the lowest order contribution shows secular growth and second, the growth is characterised by a spacetime function, $a(t)$~\cite{Kamenshchik:2020yyn, Kamenshchik:2021tjh, Miao:2021gic}.   
Recently in~\cite{Kamenshchik:2020yyn, Kamenshchik:2021tjh}, the authors proposed a renormalisation group inspired autonomous method to make a resummation of the secular effects corresponding to the one and two loop self energies of the  $\phi^4$ theory, for both massless and massive cases.  The resummed result was not only bounded at late times, but also was shown to be  in good agreement with the non-perturbative stochastic method~\cite{Starobinsky:1994bd}. We wish to apply the method of~\cite{Kamenshchik:2020yyn} below in order to resum the expressions of $\langle \phi \rangle $, \ref{l66}, \ref{l80} and attempt to obtain a non-perturbative result in the asymptotic future. We have at late times,
\begin{eqnarray}
\langle \phi \rangle = -\frac{\beta}{2^4 \times 3\pi^2}\ln^2 a + \frac{\beta \lambda}{2^7 \times 3 \pi^4}\times \frac{11\ln^4 a}{54} = -\langle \psi \rangle \quad ({\rm say})
\label{l87}
\end{eqnarray}
We change now to the cosmological time, $a= e^{Ht}$ and define a dimensionless time variable $u=Ht$. Then from \ref{l87}, and for $\beta>0$, we write
\begin{eqnarray}
\frac{d\langle\overline{ \psi} \rangle}{du} = \frac{u}{2^3\times 3 \pi^2}- \frac{ \lambda}{2^5 \times 3 \pi^4}\times \frac{11 u^3}{54}
\label{l89}
\end{eqnarray}
where we have defined the dimensionless quantity, $\langle\overline{ \psi} \rangle = \langle \psi \rangle/\beta$, for our convenience. Note that when the above equation is integrated perturbatively by breaking $\langle\overline{ \psi} \rangle$ into  ${\cal O}(\lambda^0)$ and ${\cal O}(\lambda)$ parts, we reproduce \ref{l87}. The method of~\cite{Kamenshchik:2020yyn} is then based upon increasing the domain of perturbative dependence to the non-perturbative one as follows. We note from \ref{l87} that at the leading and next to the leading order,
$$ u = 4\sqrt{3}\pi \langle \overline{\psi} \rangle^{1/2} + \frac{22\sqrt{3}\pi \lambda}{9}\langle \overline{\psi} \rangle^{3/2}+{\cal O}(\lambda^2)$$
We use the above expression to replace the dimensionless time variable $u$ in favour of $\langle\overline{ \psi} \rangle $ in \ref{l89},
\begin{eqnarray}
\frac{d\langle\overline{ \psi} \rangle}{du} = \frac{\langle\overline{ \psi} \rangle^{1/2} }{2\sqrt{3} \pi}- \frac{11 \lambda (4\sqrt{3}-1)}{108 \pi} \langle\overline{ \psi} \rangle^{3/2},
\label{l90}
\end{eqnarray}
and treat  $ \langle\overline{ \psi} \rangle$ appearing above non-perturbatively. In other words, any perturbative expansion of  $ \langle\overline{ \psi} \rangle$ in the above equation is now forbidden, thereby promoting  the same to the non-perturbative domain. Defining now a new dimenionless variable, $\langle\overline{ \psi} \rangle=\chi^2$, the above equation is rewritten as,
\begin{eqnarray}
\frac{d\chi}{du} = c_1-c_2\chi^2, 
\label{l90'}
\end{eqnarray}
where we have abbreviated,
$$c_1= \frac{1}{4\sqrt{3} \pi} \qquad c_2=  \frac{11\lambda (4\sqrt{3}-1) }{216\pi}$$
Integrating \ref{l90'} we have 
\begin{eqnarray}
\ln|\chi -\sqrt{c_1/c_2}|- \ln|\chi +\sqrt{c_1/c_2}|= -2\sqrt{c_1c_2} u
\label{l91}
\end{eqnarray}
where we have set the integration constant to zero in order 
to formally comply with the fact that both sides are consistent as $u \to 0$. However, note that  \ref{l91} is clearly not trustworthy at early times.

From \ref{l91} it is clear that as $u \to \infty$, we must have, $\chi =+\sqrt{c_1/c_2}$, which gives  
\begin{eqnarray}
\langle \phi \rangle \big\vert_{a\gg1}= -\beta\frac{c_1}{c_2}= -\frac{54\beta}{11\sqrt{3}\lambda (4\sqrt{3}-1)} \approx -\frac{0.4781\,\beta}{\lambda}
\label{l92}
\end{eqnarray}
which is negative since we have taken $\beta>0$, as expected from  \ref{f1}.   \ref{l92}, along with \ref{l80} and \ref{finalcu} are the main results of this paper. As  $\lambda \to 0$, $\langle \phi \rangle $ diverges,  reflecting the fact that the cubic potential is unbounded from below and hence must lead to instability.

Note that the expression of \ref{l92} is independent of $H$, whereas in the flat spacetime we have no such secular effect. This apparently misleading feature may correspond to the fact that once we set $u=Ht \to \infty$ in \ref{l91}, setting further $H=0$ does not make any sense. Note also that the minima of  \ref{f1},  corresponds to $\phi=-3\beta/\lambda$, which is 
approximately one order of magnitude larger than the quantum result of \ref{l92}. This should correspond to the fact that due to radiative corrections, and dynamical mass generation which we have not computed here, the shape and size of the potential must change at late times.

\section{Discussion}\label{disc}
\noindent
We have considered a massless and minimal scalar field theory with an asymmetric self interaction, $V(\phi)= \lambda \phi^4/4!+ \beta \phi^3/3!$, in the inflationary de Sitter spacetime. We have computed the renormalised two loop vacuum expectation value of the energy momentum tensor, \ref{finalcu}, owing to the cubic self interaction at the leading secular order in \ref{pot}. Using some of the results we derive here, we next compute in \ref{tadpole} the renormalised vacuum expectation value of the field operator $\phi(x)$ at one  and two loop orders (${\cal O}(\beta)$ and ${\cal O}(\lambda \beta)$), \ref{l66}, \ref{l80}. Each of these perturbative results shows monotonic growth at late cosmological times in terms of the de Sitter breaking secular logarithms. Finally, we find out a   finite, resummed non-perturbative late time expression of $\langle \phi \rangle$, \ref{l92}, using a renormalisation group inspired autonomous method proposed recently in~\cite{Kamenshchik:2020yyn}, eventually showing that the de Sitter symmetry is maintained at late times. 

 As we have discussed earlier, the chief motivation behind this study comes from the fact that such a potential can generate negative vacuum expectation value of $V(\phi)$, as is expected from  \ref{f1}, thereby reducing the value of $\Lambda$ at late times. Apart from $\langle V(\phi)\rangle$, the dynamically generated $\langle \phi \rangle$ will act as an effective  cosmological constant at late times too, while the system was assumed to be located around $\phi \sim 0$ at the initial time $t=0$. We note that  $\langle \phi \rangle$, \ref{l92}, turns out to be approximately one order of magnitude less than the position of the classical minima of the potential, $\phi=-3\beta/\lambda$. Had we used further higher order perturbative results to perform the resummation, we believe to have \ref{l92} improved. Although we do not expect it to merge with the above classical minima, as \ref{l92} should correspond to the minima of an effective potential containing the signature of the radiative 
processes. Finally, we wish to emphasise that even though there is no reason {\it a priori} to expect that we may generate large backreaction via these non-perturbative results, computing and estimating them using quantum field theory nevertheless is an important task, similar to that of the quartic self interaction.

 Accordingly, as of \ref{l2} we add in the action a term $\gamma \left(R- d(d-1)H^2 \right)\phi $, where $\gamma$ is a constant of length dimension $-1$. Since $R=12H^2$ in $d=4$, this choice guarantees that the Klein-Gordon equation does not get a source, and hence the initial state is indeed the Bunch-Davies vacuum.  As of \ref{l3} then, the leading and resummed late time contribution of this term to the energy-momentum tensor would be : $-2\gamma \Lambda \langle \phi \rangle g_{\mu\nu}$.
Accordingly, the backreaction to $\Lambda$ in the Einstein equation will be,
$$\Lambda \to \Lambda \left( 1+16\pi G \gamma \langle \phi \rangle \right)\approx\Lambda \left( 1-  8.43\times \frac{L_P^2}{L_C L_{\gamma}} \frac{\overline{\beta}}{\lambda} \right) $$
where we have used \ref{l92}, and $\overline{\beta}=\beta/H$ is dimensionless. $H^{-1}=L_C$ is the length scale of the cosmological event horizon, whereas $L_{\gamma}=\gamma^{-1}$ is the length scale associated with $\gamma$. The Hubble rate during the inflation can be estimated as~\cite{Enqvist:2017kzh}, $H =  8 \times 10^{13}\sqrt{r/0.1}\, {\rm GeV} $,
where $r< 0.03$ is the primordial tensor to scalar ratio. 
This gives the lower bound, $L_{C} = H^{-1}\gtrsim 10^{-30}{\rm m}$.  Thus we have  $L_P/L_C \lesssim {\cal O}(10^{-5})  $. Hence  the approximate shift in the inflationary $\Lambda$ is given by,
$$ \Lambda\left(1-4.1\times 10^{-9}\times \frac{\overline{\beta} L_C}{\lambda L_{\gamma}} \right)  $$
We could not unfortunately determine $L_{\gamma}$ uniquely from any physical principal. For example, if we take $L_{\gamma} \sim L_C$ and $\overline{\beta}/\lambda \sim {\cal O }(1)$, we find a shift in $\Lambda$ of ${\cal O}(10^{-8})$.  Any significant shift in $\Lambda$ seems possible only if we agree to take $L_{\gamma}$ to be as small as $L_P$ and $\overline{\beta}/\lambda$ to be as large as   ${\cal O}(10^2)$ or ${\cal O}(10^{3})$. 

Let us also check the initial backreaction as $t \to 0$ subject to these choices of $L_{\gamma}$ and $\overline{\beta}/\lambda$. We want the backreaction due to $\langle \phi \rangle$ must be small compared to $\Lambda$ then.  The leading perturbative contribution comes from \ref{l66} at initial times. Since $\langle \phi \rangle_{t\to 0} =0 $, the non-vanishing contribution to $\langle \phi \rangle$ comes from the derivatives  as of the conformal counterterm of \ref{l3}. Accordingly, we estimate the shift in $\Lambda$,
$$\Lambda \left(1+ 4.1\times 10^{-12} \times \frac{L_{C}}{L_{\gamma}} \frac{\overline \beta}{\lambda}\lambda \right)$$
 Thus we may safely ignore the initial backreaction with the choices $L_{\gamma} \sim L_P$ and   $\overline{\beta}/\lambda$ as large as $\sim {\cal O}(10^3)$, with any reasonable $\lambda \sim {\cal O}(1)$. However  we are not sure as of now, whether the above  choices will be acceptable, when tested on the basis of some other physical constraints, e.g. the cosmological correlation functions.
 
Before we end, perhaps it is worthwhile comparing our result with the prediction of the stochastic method~\cite{Starobinsky}. This is essentially a late time infrared effective formalism, where the field $\phi$ is expected to be at the minimum of the quantum corrected $V(\phi)$. In our field theoretic analysis however, we assumed the field to be located around $\phi\sim 0$ in \ref{f1}, {\it initially} ($t\to 0$).  As we have mentioned earlier, this assumption is necessary to assert that we begin with a large inflationary value of $\Lambda$ at $t=0$. This also guarantees that the perturbation theory is valid initially. As time goes on, the field tends to roll down towards the minimum of $V(\phi)$, which gets radiatively corrected due to the quantum fluctuations. This leads to the dynamically generated $\langle \phi \rangle$ at late times, which may partially screen the inflationary $\Lambda$.   Since  $V(\phi)$ is bounded from below, following~\cite{Starobinsky}, we have
\be
\langle \overline{\phi} \rangle_{\rm stochastic}=\frac{\int_{-\infty}^{\infty} d\overline{\phi}\, \overline{\phi} e^{-8\pi^2 V(\overline{\phi})/3}}{\int_{-\infty}^{\infty} d\overline{\phi}'\, e^{-8\pi^2 V(\overline{\phi}')/3}  }
\label{stoch}
\ee  
  where $\overline{\phi}=\phi/H $. The above expression can easily be evaluated numerically in MATHEMATICA.  For example, for $\lambda=0.1$ and $\overline{\beta}/\lambda \sim 10^{-2}$, the above yields $\langle \overline{\phi} \rangle \sim -0.01$, which is approximately two times larger than the quantum field theory prediction,~\ref{l92}. For $\lambda=0.1$ and $\overline{\beta}=0.01$, the above gives $\langle \overline{\phi} \rangle \sim -0.1$, also approximately  twice the field theory prediction, $-0.05$.
    For $\lambda=\overline{\beta}=0.1$, the above gives $\langle \overline{\phi} \rangle \sim 2.57$, approximately $5$ times larger than that of~\ref{l92}. For $\lambda = 0.1$ and $\overline{\beta}\sim 10^2$, the stochastic formalism suggests $\langle \overline{\phi} \rangle \sim - 3043$, approximately six times larger than that of  the field theory prediction.  Note also that \ref{l92} depends only upon the ratio $\overline{\beta}/\lambda$, whereas \ref{stoch} will clearly depend upon that ratio as well as the $\lambda$-value. Apparently, such mismatches may correspond to the choice of our initial condition. As of now, it is not clear to us how such initial condition may be translated into the stochastic formalism. However, note also that our resummation, \ref{resum}, essentially involves two coupling constants and the mismatch between the results of the two formalisms grows with increasing $\overline{\beta}$. Hence it might also be possible that such resummation essentially requires more higher order terms, in particular, involving {\it both} $\lambda$ and $\beta$.  We hope to investigate this issue in detail in our future works.

There are a couple of things which  perhaps warrant further attention about the model of the potential we have studied here. For example,  computation of the non-perturbative vacuum expectation value of $V(\phi)$ and to see how it compares with the above result. While computations at ${\cal O}(\lambda)$ has been done earlier~\cite{Onemli:2002hr} and we have computed it at ${\cal O}(\beta^2)$, \ref{finalcu}, intuitively it seems that  any resummation technique will be more efficient if we go upto the order of perturbation theory where $\lambda$ and $\beta$ are overlapping, as mentioned earlier. This will require three loop computations at ${\cal O}(\lambda^2)$ and ${\cal O}(\lambda \beta^2)$. A non-perturbative computation of the dynamical generation of mass seems to be interesting as well. Finally, investigating this theory via the stochastic formalism, to compare it with the field theoretic results and finding the effects of the  quartic-cubic coupling  on various cosmological correlation functions will be interesting. We refer our reader to~\cite{Markkanen:2020bfc} for a recent interesting discussion  on an asymmetric potential  different from the above, in the context of cosmological correlation functions. We believe the above problems are not only interesting technically, but also might predict something  interesting physically. We shall come back to them in our future publications.

\section*{Acknowledgements}
This research was partially supported by the ISIRD grant 9-289/2017/IITRPR/704. I would like to sincerely acknowledge late Prof Theodore N Tomaras for discussions  and encouragements on de Sitter quantum field theory in numerous occasions   over the past  years. I thank  Nitin Joshi and Moutushi for their help to draw the diagrams. I would also like to thank anonymous referee for careful critical reading of the manuscript and for useful comments.
\bigskip
\appendix
\labelformat{section}{Appendix #1} 
\section{Computations pertaining \ref{lcphi4} }\label{A}

In this Appendix we shall outline the derivation~\cite{Onemli:2002hr} of the first integral of \ref{k3}, coming from \ref{l21}. The integral reads, after using \ref{l8} and recalling $d=4-\e$,
\begin{eqnarray}
{ I}_{\mu\nu}=  -\frac{i\lambda H^{8-3\e}}{2^{7-\e} \pi^{6-3\e/2}}\frac{\Gamma(2-\e)\Gamma^2(2-\e/2)}{\Gamma(1-\e/2)}{ \overline I}_{\mu\nu}
\label{l22'}
\end{eqnarray}
where
\begin{eqnarray}
{\overline I}_{\mu\nu} = 4a^{-4+\e} H^{-6+2\e}  \left(\eta_{\mu \alpha} \eta_{\nu \beta}-\frac12 \eta_{\mu\nu} \eta_{\alpha\beta}\right) \,\int d^d x' a'^2 \ln a' \Delta x^{\alpha} \Delta x^{\beta} \left(\frac{1}{\Delta x_{++}^{8-2\e}}-\frac{1}{\Delta x_{+-}^{8-2\e}} \right)
\label{l22}
\end{eqnarray}
Some useful differential identities are given by,
\begin{eqnarray}
&&\p_{\alpha}\p_{\beta}\frac{1}{\Delta x^{4-2\e}} =-(4-2\e) \left[\frac{\eta_{\alpha\beta}}{\Delta x^{6-2\e}} - \frac{(6-2\e) \Delta x_{\alpha}  \Delta x_{\beta}}{\Delta x^{8-2\e} } \right], \qquad
\p^2\frac{1}{\Delta x^{4-2\e}} =  \frac{2(2-\e)^2}{\Delta x^{6-2\e}} \nonumber\\
&&  \p_{\alpha}\p_{\beta}\frac{1}{\Delta x^{2-2\e}} = -2(1-\e)\left[ \frac{\eta_{\alpha \beta}}{\Delta x^{4-2\e}} - \frac{2(2-\e)\Delta x_{\alpha} \Delta x_{\beta}}{\Delta x^{6-2\e}}\right], \qquad 
\p^2 \frac{1}{\Delta x^{2-2\e}} = -\frac{2\e (1-\e)}{\Delta x^{4-2\e}}
\label{l23}
\end{eqnarray}
Using the above relations, we have,
\begin{eqnarray}
&& \frac{\left( \delta_{\beta 0}\Delta x_{\alpha}+\delta_{\alpha 0}\Delta x_{\beta} \right)}{\Delta x^{6-2\e} }= \frac{1}{4\e(2-\e)(1-\e)}\left( \delta_{\beta 0}\p_{\alpha}+\delta_{\alpha 0}\p_{\beta}\right)\p^2 \frac{1}{\Delta x^{2-2\e} }\nonumber\\
&&\frac{\Delta x_{\alpha} \Delta x_{\beta}}{\Delta x^{6-2\e}}= \frac{1}{4(1-\e)(2-\e)} \left(\p_{\alpha}\p_{\beta} -\frac{\eta_{\alpha\beta}}{\e}\p^2\right)\frac{1}{\Delta x^{2-2\e}}\nonumber\\
&&\frac{\Delta x_{\alpha} \Delta x_{\beta}}{\Delta x^{8-2\e}} = -\frac{1}{8\e(1-\e)(2-\e)(3-\e) } \left( \p_{\alpha} \p_{\beta} + \frac{\eta_{\alpha \beta}}{2-\e} \p^2\right)\p^2 \frac{1}{\Delta x^{2-2\e}}
\label{l24}
\end{eqnarray}
Substituting this into \ref{l22}, we have
\begin{eqnarray}
{\overline I}_{\mu\nu} = -\frac{a^{-4+\e} H^{-6+2\e} }{2\e(1-\e)(2-\e)(3-\e)} (\p_{\mu}\p_{\nu} -\eta_{\mu \nu} \p^2)\int d^d x' a'^2 \ln a' \p^2 \left( \frac{1}{\Delta x_{++}^{2-2\e}}-\frac{1}{\Delta x_{+-}^{2-2\e}} \right)
\label{l25}
\end{eqnarray}

Using now \ref{l8}  we get
\begin{eqnarray}
\p^2 \frac{1}{\Delta x_{++}^{2-\e}}&& = - \frac{(2-\e)(d -i\e \p_0^2 |\eta -\eta'|)}{[(x-x')^2 + 2i \e |\eta -\eta'|+\e^2]^{2-\e/2}} + \frac{(2-\e)d\left((x-x')^2 + 2i\e (x-x')^{\mu}\p_{\mu} |\eta -\eta'|\right)}{[(x-x')^2 + 2i \e |\eta -\eta'|+\e^2]^{3-\e/2}}\nonumber\\
&&=\frac{i \e (2-\e) \p_0^2 |\eta -\eta'|}{[(x-x')^2 + 2i \e |\eta -\eta'|+\e^2]^{2-\e/2}} +{\cal O} (\e)
\label{l26}
\end{eqnarray}
Now $\p_0 |\eta -\eta'| ={\rm Sgn} (\eta -\eta')$ and hence $\p_0^2 |\eta -\eta'| =2\delta (\eta -\eta')$, can be proven by integration. Thus
\be
\p^2 \frac{1}{\Delta x_{++}^{2-\e}} = \frac{2i\e(2-\e)\delta (\eta-\eta')}{[(\vec{x}-\vec{x}')^2 + \e^2]^{2-\e/2}}\to \frac{4i \pi^{2-\e/2} }{\Gamma(1-\e/2)}\delta^d(x-x')
\label{l27}
\ee
and also
$$\p^2 \frac{1}{\Delta x_{+-}^{2-\e}} = 0$$
We have 
$$\p^2 \frac{1}{\Delta x_{+-}^{2-2\e}} = \p^2\left(\frac{1}{\Delta x_{+-}^{2-2\e}} -\frac{\mu^{-\e}}{\Delta x_{+-}^{2-\e}} \right) $$
where the second term vanishes and $\mu$ is a mass scale introduced to maintain dimensional consistency. The UV finite term appearing above can be rewritten as 
\be
\p^2 \frac{1}{\Delta x_{+-}^{2-2\e}} =\e\, \p^2 \frac{\ln\, \mu^2 \Delta x_{+-}^{2}}{2\Delta x_{+-}^{2}} 
\label{l28}
\ee
Likewise, using \ref{l27} we have
\be
\p^2 \frac{1}{\Delta x_{++}^{2-2\e}} =  \p^2 \left( \frac{1}{\Delta x_{++}^{2-2\e}} - \frac{\mu^{-\e}}{\Delta x_{++}^{2-\e}}\right)+ \frac{4i \mu^{-\e}\,\pi^{2-\e/2} }{\Gamma(1-\e/2)}\delta^d(x-x')= \e\, \p^2 \frac{\ln\, \mu^2 \Delta x_{++}^{2}}{2\Delta x_{++}^{2}}+ \frac{4i \mu^{-\e}\,\pi^{2-\e/2} }{\Gamma(1-\e/2)}\delta^d(x-x')
\label{l29}
\ee
Substituting \ref{l27}, \ref{l29} into \ref{l25}, we obtain
\begin{eqnarray}
{\overline I}_{\mu\nu}= &&-\frac{2i \mu^{-\e} a^{\e} H^{-4+2\e} \pi^{2-\e/2}}{\e (1-\e)(2-\e)(3-\e)\Gamma(1-\e/2)} (\delta_{\mu}{}^0 \delta_{\nu}{}^0 +\eta_{\mu\nu})(6\ln a +5)\nonumber\\&&+\frac{a^{-4+\e} H^{-6+2\e} }{2^2(1-\e)(2-\e)(3-\e)} (\delta_{\mu}{}^0 \delta_{\nu}{}^0 +\eta_{\mu\nu} )\p_0^4\int d^d x' a'^2 \ln a' \, \left(\frac{\ln\, \mu^2 \Delta x_{++}^{2}}{\Delta x_{++}^{2}} -\frac{\ln\, \mu^2 \Delta x_{+-}^{2}}{\Delta x_{+-}^{2}} \right)
\label{l30}
\end{eqnarray}
where we have used the fact that the integral does not depend upon the spatial position. We also have,
\begin{eqnarray}
\p^2 \ln^2 \Delta x^2= 8 \left( \frac{\ln \Delta x^2}{\Delta x^2}+ \frac{1}{\Delta x^2}\right), \qquad 
\p^2 \ln \Delta x^2 = \frac{4}{\Delta x^2}
\label{l31}
\end{eqnarray}
Thus,
\be
\frac{\ln \Delta x^2}{\Delta x^2} =\frac18\, \p^2 (\ln^2 \Delta x^2 -2 \ln \Delta x^2 ),
\label{l31'}
\ee
Substituting the above  into \ref{l30}, the integral becomes
\begin{eqnarray}
&&\p_0^4\int d^d x' a'^2 \ln a' \, \left(\frac{\ln\, \mu^2 \Delta x_{++}^{2}}{\Delta x_{++}^{2}} -\frac{\ln\, \mu^2 \Delta x_{+-}^{2}}{\Delta x_{+-}^{2}} \right) \nonumber\\&&= -\frac{\p_0^6}{2^3}\int d^d x' a'^2 \ln a' \,[(\ln^2 \mu^2\Delta x^2_{++}- \ln^2 \mu^2\Delta x^2_{+-}) -2 (\ln \mu^2\Delta x^2_{++}- \ln \mu^2 \Delta x^2_{+-}) ]
\label{l32}
\end{eqnarray}
Breaking the logarithm into amplitude and argument  part, and writing $|\vec{x}-\vec{x}'| =r$, we have 
\begin{eqnarray}
\ln \mu^2 \Delta x_{++}^2 = \ln \mu^2 (\Delta \eta^2 -r^2) + i\pi \theta(\Delta \eta^2 -r^2), \qquad \ln \mu^2 \Delta x_{+-}^2 = \ln \mu^2 (\Delta \eta^2 -r^2) - i\pi \theta(\Delta \eta^2 -r^2) \nonumber\\
(\ln^2 \mu^2\Delta x^2_{++}- \ln^2 \mu^2\Delta x^2_{+-}) -2 (\ln \mu^2\Delta x^2_{++}- \ln \mu^2 \Delta x^2_{+-}) = 4\pi i \left(\ln \mu^2 (\Delta \eta^2 -r^2) -1\right)\theta(\Delta \eta^2 -r^2)
\label{l33}
\end{eqnarray}
Substituting this into \ref{l32}, and performing the angular integration, we have (Mathematica was used for the integrations and some of the  summations)
\begin{eqnarray}
&&-2\pi^2 i\,\p_0^6\int_{-1/H}^{\eta} d \eta' a'^2 \ln a' \, \int_{0}^{\Delta \eta} dr r^2 (\ln \mu^2(\Delta \eta^2 -r^2)-1 )\nonumber\\ && = -2\pi^2 i\,\p_0^6\int_{-1/H}^{\eta} d \eta' a'^2 \ln a' \,(\Delta \eta)^3 \left(\frac23 \ln 2\mu \Delta \eta -\frac{11}{9} \right)
\nonumber\\ && = -\frac{8\pi^2 i}{H}\,\p_0^3\left[ \ln \frac{2\mu}{H}\left(1-a+a\ln a\right)+ (2-2a+2a \ln a-a \ln^2 a)-\sum_{n=1}^{\infty} \frac{1}{n(n+1)^2}\left(\frac{1}{a^n}-a +(n+1)a \ln a\right) \right] \nonumber\\
&&=-8i\pi^2H^2 a^4\left[ \ln \frac{2\mu}{H}(6\ln a +5) -6\ln^2 a-16\ln a -1-\pi^2  + \sum_{n=1}^{\infty} \frac{(n-1)(n-2)}{(n+1)^2} a^{-n-1}\right]
 \label{l34}
\end{eqnarray}
where we also have used $d\eta = da/Ha^2$, and 
$$\int_{0}^{\Delta \eta} dr r^2 \left(\ln \mu^2 (\Delta \eta^2 -r^2)-1 \right)= \frac{(\Delta \eta)^3}{3}\left(2 \ln 2\mu \Delta \eta -\frac{11}{3} \right), \qquad \p_0^3\left[ \frac{(\Delta \eta)^3}{3} \left( 2\ln 2\mu \Delta \eta -\frac{11}{3}\right)\right]= 4 \ln 2\mu \Delta \eta$$
Combining now \ref{l34} with \ref{l30} and \ref{l22'}, the first integral of \ref{l21} becomes
\begin{eqnarray}
I_{\mu\nu}&& =-\frac{i \lambda a^2 H^{8-3\e} \Gamma(2-\e/2)\Gamma(3-\e)}{2^{8-\e} \pi^{6-3\e/2}} \,\overline{I}^1_{\mu\nu} \nonumber\\
&&=\frac{i \lambda a^2 H^{8-3\e} \Gamma(2-\e/2)\Gamma(3-\e)}{2^{8-\e} \pi^{6-3\e/2}} \left[ \frac{2i \mu^{-\e} a^{\e} H^{-4+2\e} \pi^{2-\e/2}}{\e (1-\e)(2-\e)(3-\e)\Gamma(1-\e/2)} (\delta_{\mu}{}^0 \delta_{\nu}{}^0 +\eta_{\mu\nu})(6\ln a +5)\right. \nonumber\\&& \left.  +\frac{ i\pi^2 H^{-4} }{3} (\delta_{\mu}{}^0 \delta_{\nu}{}^0 +\eta_{\mu\nu} )\left( \ln \frac{2\mu}{H}(6\ln a +5) -6\ln^2 a-16\ln a -1-\pi^2  + \sum_{n=1}^{\infty} \frac{(n-1)(n-2)}{(n+1)^2} a^{-n-1}\right)  \right] \nonumber\\
&&= -    (\delta_{\mu}{}^0 \delta_{\nu}{}^0 +\eta_{\mu\nu})\frac{\lambda a^2 H^4}{2^6\pi^4}  \left[\ln a \left(\frac{\zeta}{\e}  -\frac32+\ln \frac{2\mu}{H}  \right)+ \frac56\left(\frac{\zeta}{\e}+ \ln \frac{2\mu}{H}\right) +\frac19 - \frac{\pi^2}{6}+\sum_{n=3}^{\infty}\frac{(n-1)(n-2)}{6(n+1)^2} a^{-n-1}    \right]              \nonumber\\                       
 \label{l35}
\end{eqnarray}
where we have abbreviated, 
\be
\zeta = \left(\frac{2\pi}{\mu H} \right)^{\e} \left(1-\frac{\e}{2}\right)\Gamma(1-\e)
\label{l36}
\ee 
The remaining five integrals of \ref{k3} can be evaluated in a likewise manner. Combining all the results, \ref{l41} follows.

\section{One loop tadpole and mass renormalisation counterterms for the cubic self interaction }\label{B}

Let us first determine the one loop tadpole counterterm  $\alpha$ in \ref{l3}, corresponding to the first row of \ref{f4}. Using \ref{l13}, this contribution equals
\begin{eqnarray}
-i a^d \left(\frac{\beta}{2} i \Delta(x,x)+\alpha \right) = - i a^d \left[\frac{\beta H^2}{2^4\pi^2} \left(\frac{H^2}{4\pi} \right)^{-\frac{\e}{2}} \frac{\Gamma(3-\e)}{\Gamma(2-\frac{\e}{2})} \left(\frac{1}{\e}+ \ln a \right) +\alpha \right]
\label{b1}
\end{eqnarray} 
which gives
\begin{eqnarray}
\alpha = - \frac{\beta H^{2-\e}}{2^{4-\e} \pi^{2-\e/2}} \frac{\Gamma(3-\e)}{\Gamma(2-\e/2)}\frac{1}{\e}
\label{b2}
\end{eqnarray} 

Likewise, the contribution for the one loop self energy equals for the $++$ interaction  (the second row of \ref{f4}),
\begin{eqnarray}
-iM^2_{++,\,\phi^3} (x,x')= -\frac{\beta^2}{2} a^d a'^d (i \Delta_{++} (x,x'))^2 - ia^d \delta m^2_{\beta} \delta^d(x-x')
\label{b2'}
\end{eqnarray} 
Using \ref{l7}, \ref{l8}, \ref{l8'}  we have
\begin{eqnarray}
(i \Delta_{++} (x,x'))^2= \frac{(aa')^{-2+\e} \Gamma^2(1-\e/2)}{2^4 \pi^{4-\e}} \frac{1}{\Delta x_{++}^{4-2\e}} +\frac{H^4}{2^6 \pi^4} \ln^2 \frac{\sqrt{e} H^2 \Delta x_{++}^2}{4} -\frac{H^2 (aa')^{-1}}{2^4 \pi^4} \frac{\ln \frac{\sqrt{e} H^2 \Delta x_{++}^2}{4}}{\Delta x_{++}^2}
\label{b3}
\end{eqnarray} 

Only the first term is ultraviolet divergent. Using  \ref{l23}, \ref{l29}, we find from \ref{b2} 
\begin{eqnarray}
-iM^2_{++,\,\phi^3} = i \left(\frac{ \beta^2 \mu^{-\e} a^4 \Gamma(1-\e/2)}{2^4 \pi^{2-\e/2} \e (1-\e)} -a^d \delta m^2_{\beta}   \right)\delta^d(x-x')\,+\,\,{\rm UV~finite~terms }
\label{b4}
\end{eqnarray} 
Using $a^4= a^d (1+ \e \ln a)$, we find
\begin{eqnarray}
\delta m^2_{\beta} = \frac{ \beta^2 \mu^{-\e} \Gamma(1-\e/2)}{2^4 \pi^{2-\e/2} \e (1-\e)}
\label{b5}
\end{eqnarray} 
%

\section{Computations pertaining \ref{pot} }\label{C}
Let us first evaluate the integral,
\be
\int  d^d x' a'^d \ \ln a' \left( i \Delta_{++}(x,x')- i \Delta_{+-}(x,x')  \right)
\label{c10}
\ee
As is evident from the detail given at the beginning of \ref{A}, the above integral is not ultraviolet divergent. Using then \ref{l7}, \ref{l8}, we write
\begin{eqnarray}
&&i \Delta_{++}(x,x')- i \Delta_{+-}(x,x') = \frac{1 }{4 \pi^2 aa'} \left(\frac{1}{\Delta x_{++}^2} -\frac{1}{\Delta x_{+-}^2} \right) -\frac{H^{2} }{2^{3} \pi^2} \left(\ln H^2\Delta x_{++}^2 -\ln H^2\Delta x_{+-}^2 \right) +{\cal O}(\e)
\qquad
\label{c11}
\end{eqnarray} 
In the above expression $C(x,x')$ in \ref{l7} is not relevant, for it contributes at ${\cal O}(\e^2)$. We substitute the above into \ref{c10}.  Using \ref{l31}, \ref{l33}  we find respectively for the above two terms,
\begin{eqnarray}
-\frac{1}{2^4 \pi^{2} a} \p_0^{2} \int d^4 x' a'^3 \ln a' \left( \ln H^2\Delta x_{++}^2 -\ln H^2\Delta x_{+-}^2 \right) 
=-\frac{i}{H^2} \left(\frac{1}{a}-\frac{3 }{4}-\frac{1}{4 a^{2}}+\frac{1}{2} \ln a\right) + {\cal O}(\e) 
\label{c12}
\end{eqnarray} 
\begin{eqnarray}
 \frac{H^{2}  }{2^{3} \pi^{2}}\int  d^d x' a'^d  \ln a'  \left(\ln H^2\Delta x_{++}^2 -\ln H^2\Delta x_{+-}^2 \right) =   \frac{i }{3 H^2}\left(\frac{85}{36} -\frac{3}{a}+\frac{3}{4 a^{2}}-\frac{1}{9 a^{3}}+\frac{1}{2} \ln^2 a-\frac{11 \ln a}{6}\right)+{\cal O}(\e) 
\label{c13}
\end{eqnarray} 

Combining now \ref{c12}, \ref{c13},  the integral   \ref{c10} becomes,  
\begin{eqnarray}
-  \frac{i}{3H^2 }\left( \frac19-\frac{1}{9a^3}-\frac{\ln a}{3}+\frac{\ln^2 a}{2}\right)+{\cal O}(\e)
\label{c16}
\end{eqnarray} 
\bigskip

We shall now find out the integral of the cube of the propagator, the cube being computed in~\cite{Brunier:2004sb} earlier. Using \ref{l7}, \ref{l8}, \ref{l8'}, we write,
\begin{eqnarray}
&&(i\Delta_{++}(x,x'))^3-(i\Delta_{+-}(x,x'))^3 = \frac{\Gamma^3(1-\e/2) (aa')^{-3+3\e/2}}{2^6 \pi^{6-3\e/2}} \left( \frac{1}{\Delta x_{++}^{6-3\e}}- \frac{1}{\Delta x_{+-}^{6-3\e}}\right)\nonumber\\&&+ \frac{3H^{2-\e} \Gamma(1-\e/2) \Gamma(2-\e)}{2^{7-\e} \pi^{6-3\e/2}} (aa')^{-2+\e}\left[- \frac{2\Gamma(3-\e/2) \Gamma(2-\e/2)\,\left(\frac{aa'H^2}{4}\right)^{\e/2}}{\e\,\Gamma(3-\e)}\left( \frac{1}{\Delta x_{++}^{4-3\e}}- \frac{1}{\Delta x_{+-}^{4-3\e}} \right)\right. \nonumber\\ &&\left.+ \left(\frac{2}{\e} + \ln aa' \right)\left( \frac{1}{\Delta x_{++}^{4-2\e}} -\frac{1}{\Delta x_{+-}^{4-2\e}} \right) \right] + \frac{3H^4}{2^8 \pi^6 aa'}\left( \frac{\ln^2 \frac{\sqrt{e} H^2 \Delta x_{++}^2}{4}}{\Delta x_{++}^2} - \frac{\ln^2 \frac{\sqrt{e} H^2 \Delta x_{+-}^2}{4}}{\Delta x_{+-}^2}\right) \nonumber\\ &&-\frac{H^6}{2^9 \pi^6}  \left(\ln^3  \frac{\sqrt{e} H^2 \Delta x_{++}^2}{4}- \ln^3  \frac{\sqrt{e} H^2 \Delta x_{+-}^2}{4} \right)
\label{c1}
\end{eqnarray} 

We use 
$$\frac{1}{\Delta x^{4-3\e}_{+ \pm} } = -\frac{1}{2\e(2-3\e)} \p^2 \frac{1}{\Delta x^{2-3\e}_{+\pm}}, \qquad \frac{1}{\Delta x^{6-3\e}_{+ \pm} } = - \frac{1}{4 \e (1-\e) (2-3\e) (4-3\e)} \p^4 \frac{1}{\Delta x_{+\pm}^{2-3\e}}$$
and  \ref{l23}, \ref{l27} and also the method described at the beginning of \ref{A}, to find
\begin{eqnarray}
&&\frac{1}{\Delta x_{++}^{6-3\e}}- \frac{1}{\Delta x_{+-}^{6-3\e}} = - \frac{i \mu^{-2\e}\pi^{2-\e/2}}{\e (1-\e)(2-3\e)(4-3\e) \Gamma(1-\e/2)} \p^2 \delta^d(x-x') - \frac{1}{2^5} \p^4 \left(\frac{\ln \mu^2 \Delta x_{++}^2}{\Delta x_{++}^2}-\frac{\ln \mu^2 \Delta x_{+-}^2}{\Delta x_{+-}^2} \right)+{\cal O}(\e)\nonumber\\
&&\frac{1}{\Delta x_{++}^{4-3\e}}- \frac{1}{\Delta x_{+-}^{4-3\e}}= - \frac{2i \mu^{-2\e} \pi^{2-\e/2}}{\e(2-3\e) \Gamma(1-\e/2)} \delta^d(x-x')-\frac{\mu^{-3\e}}{2(2-3\e)} \p^2\left[ \left(  \frac{\ln \mu^2 \Delta x_{++}^2}{\Delta x_{++}^2}-\frac{\ln \mu^2 \Delta x_{+-}^2}{\Delta x_{+-}^2}\right)\right.\nonumber\\ &&\left.+\e \left( \frac{\ln^2 \mu^2 \Delta x_{++}^2}{\Delta x_{++}^2}-\frac{\ln^2 \mu^2 \Delta x_{+-}^2}{\Delta x_{+-}^2} \right)\right]\nonumber\\
&&\frac{1}{\Delta x_{++}^{4-2\e}}- \frac{1}{\Delta x_{+-}^{4-2\e}}= - \frac{2i \mu^{-\e} \pi^{2-\e/2}}{\e(1-\e)\Gamma(1-\e/2)} \delta^d (x-x') -\frac{\mu^{-2\e}}{4(1-\e)} \p^2 \left[ \left(  \frac{\ln \mu^2 \Delta x_{++}^2}{\Delta x_{++}^2}-\frac{\ln \mu^2 \Delta x_{+-}^2}{\Delta x_{+-}^2}\right)\right.\nonumber\\ &&\left.+\frac{3\e}{4} \left( \frac{\ln^2 \mu^2 \Delta x_{++}^2}{\Delta x_{++}^2}-\frac{\ln^2 \mu^2 \Delta x_{+-}^2}{\Delta x_{+-}^2} \right)\right]
\label{c2}
\end{eqnarray}

This gives,
\begin{eqnarray}
&&(i\Delta_{++}(x,x'))^3-(i\Delta_{+-}(x,x'))^3=-\frac{i \mu^{-2\e} \Gamma^2(1-\e/2) (aa')^{-3+3\e/2}}{2^9 \pi^{4-\e}(1-3\e/2)(1-\e)(1-3\e/4)\e}\p^2 \delta^d(x-x') + \frac{3i\mu^{-\e} H^{2-\e} a^{-4+2\e}}{2^{6-\e}\pi^{4-\e}}\times \nonumber\\
&&\left[\frac{(1-\e/2)(1-\e/4) \Gamma^2(1-\e/2)}{(1-3\e/2)\e^2}\left(\frac{Ha}{2\mu} \right)^{\e}-\frac{2\Gamma(1-\e)}{\e}\left(\frac{1}{\e}+\ln a \right) \right]\delta^d(x-x')\nonumber\\
&&- \frac{(aa')^{-3}}{2^{11}\pi^6}\p^4 \left(\frac{\ln \mu^2 \Delta x_{++}^2}{\Delta x_{++}^2} -\frac{\ln \mu^2 \Delta x_{+-}^2}{\Delta x_{+-}^2}\right)+\frac{3H^2(aa')^{-2}}{2^{8}\pi^6} \ln \left(\frac{H e^{3/4}}{2\mu}\right)\p^2 \left(\frac{\ln \mu^2 \Delta x_{++}^2}{\Delta x_{++}^2} -\frac{\ln \mu^2 \Delta x_{+-}^2}{\Delta x_{+-}^2}\right)\nonumber\\
&& + \frac{3H^2(aa')^{-2}}{2^{10}\pi^6} \p^2 \left(\frac{\ln^2 \mu^2 \Delta x_{++}^2}{\Delta x_{++}^2} -\frac{\ln^2 \mu^2 \Delta x_{+-}^2}{\Delta x_{+-}^2}\right)
+  \frac{3H^4}{2^8 \pi^6 aa'}\left( \frac{\ln^2 \frac{\sqrt{e} H^2 \Delta x_{++}^2}{4}}{\Delta x_{++}^2} - \frac{\ln^2 \frac{\sqrt{e} H^2 \Delta x_{+-}^2}{4}}{\Delta x_{+-}^2}\right)\nonumber\\&& -\frac{H^6}{2^9 \pi^6}  \left(\ln^3  \frac{\sqrt{e} H^2 \Delta x_{++}^2}{4}- \ln^3  \frac{\sqrt{e} H^2 \Delta x_{+-}^2}{4} \right)
\label{c3}
\end{eqnarray}
We substitute the above into \ref{l53}. We first find out the contribution of the first two terms on the right hand side plus the mass counterterm, giving
\begin{eqnarray}
&& - \frac{\beta^2 H^2 \mu^{-2\e} \Gamma^2(1-\e/2) (1+\e/2) (1+\e/4)}{3\times 2^9 \pi^{4-\e} (1-3\e/2)(1-\e)(1-3\e/4) \e} +\frac{\beta^2 \mu^{-\e}H^{2-\e}\Gamma(1-\e)}{2^{7-\e} \pi^{4-\e} \e^2} \left(1- \left(\frac{2\mu}{H} \right)^{-\e}\frac{(1-\e/4)\Gamma(2-\e/2)}{(1-3\e/2)}\right) \nonumber\\
&&+\frac{\beta^2\mu^{-\e}H^{2-\e}}{2^{6-\e}\pi^{4-\e}\e}\,\ln a \left(\frac{3\Gamma(1-\e)}{2}  -  \left(\frac{2\mu}{H} \right)^{-\e}\frac{\Gamma(1-\e/2)\Gamma(2-\e/2)(1-\e/4)}{(1-3\e/2)}\right)+ \frac{\beta^2 H^2}{2^7\pi^4}\left(\ln^2 a -\frac16 \ln a \right) \nonumber\\
\label{c4}
\end{eqnarray}
The   integrals corresponding to the third and fourth terms of \ref{c3} respectively equals,
\begin{eqnarray}
\begin{split}
&& - \frac{\beta^2 H^2 }{3\times 2^9 \pi^4 } \left[\left(2\ln \frac{2\mu}{H}-3 \right)-2\ln a -\sum \frac{(n-1)(n-2)}{n} a^{-n} \right], \\
&&-\frac{\beta^2 H^2}{2^6\pi^4} \ln \left(\frac{2\mu e^{-3/4}}{H}\right) \left[\ln \frac{2\mu}{H }-1-\ln a - \sum \frac{a^{-n-1}}{n+1} \right]
\end{split}
\label{c5}
\end{eqnarray}
In order to evaluate the fifth and the sixth integrals corresponding to \ref{c3}, we first note that
\begin{eqnarray}
&&\frac{\ln^2 \mu^2 \Delta x_{++}^2}{\Delta x_{++}^2} -\frac{\ln^2 \mu^2 \Delta x_{+-}^2}{\Delta x_{+-}^2}= \frac{1}{12}\p^2\left(\ln^3 \mu^2\Delta x_{++}^2-\ln^3\mu^2 \Delta x_{+-}^2 \right),\nonumber\\ && -\frac14 \p^2 \left[ \left(\ln^2 \mu^2\Delta x_{++}^2- \ln^2\mu^2\Delta x_{+-}^2 \right) -2\left(\ln \mu^2 \Delta x_{++}^2- \ln \mu^2 \Delta x_{+-}^2  \right)\right]
\nonumber\\ && =\frac{i\pi}{6}\p^2 \left(3\ln^2\mu^2(\Delta \eta^2-r^2)\theta(\Delta \eta^2 -r^2)-\pi^2 \theta^3(\Delta \eta^2 -r^2) \right)- i\pi \p^2\left(\left(\ln \mu^2 (\Delta \eta^2-r^2)-1 \right)\theta(\Delta \eta^2 -r^2)\right)
\nonumber\\
\label{c6}
\end{eqnarray}
and also,
$$ \frac{i\pi}{6}\p^2 \left(3\ln^2\mu^2(\Delta \eta^2-r^2)-\pi^2  \right)- i\pi \p^2\left(\left(\ln \mu^2 (\Delta \eta^2-r^2)-1 \right)\right) = \frac{i\pi}{2}\p^2 \left[\ln^2 \frac{\mu^2}{e}(\Delta \eta^2-r^2)+1-\frac{\pi^2}{3} \right]$$
and
$$\int_{0}^{\Delta \eta} dr r^2 \ln^2 \mu^2(\Delta \eta^2 -r^2)= (\Delta \eta)^3 \left(\frac{104}{27} -\frac{\pi^2}{9} -\frac{32}{9} \ln 2 \mu \Delta \eta +\frac43 \ln^2 2\mu \Delta \eta\right) $$
Using these, the fifth integral becomes,
\begin{eqnarray}
&&- \frac{\beta^2 H^2 a^{-2}}{2^{10}\pi^4} \frac{\p_0^4}{H^4} \int_{1}^a \frac{da'}{a'^3}\left(1-\frac{a'}{a}\right)^3\left[\left(\frac{113}{27} -\frac{2\pi^2}{9} -\frac{32}{9}\ln\frac{2\mu}{\sqrt{e}H} +\frac43 \ln^2\frac{2\mu}{\sqrt{e}H} \right)   \right. \nonumber\\ &&\left.          
+ \left(\frac{32}{9}-\frac83 \ln\frac{2\mu}{\sqrt{e}H} \right) \left(\ln a' +\sum \frac{a'^n}{n a^n}\right) +\frac43 \ln^2a' +\frac43 \sum \frac{a'^{m+n}}{mn\, a^{m+n}} + \frac83 \ln a' \sum \frac{a'^n}{na^n} \right]\nonumber\\
&&=- \frac{\beta^2 H^2 a^{-2}}{2^{10}\pi^4} \frac{\p_0^4}{H^4} \left[\left(\frac{113}{27} -\frac{2\pi^2}{9} -\frac{32}{9}\ln\frac{2\mu}{\sqrt{e}H} +\frac43 \ln^2\frac{2\mu}{\sqrt{e}H} \right) \left(\frac{1}{2}+\frac{1}{a^3}+\frac{3}{2 a^2}+\frac{3 \ln a}{a^2}-\frac{3}{a}\right) \right. \nonumber\\ && \left.+ \left(\frac{32}{9}-\frac83 \ln\frac{2\mu}{\sqrt{e}H} \right) \left[\left(\frac14-\frac{3}{a} +\frac{15}{4 a^2}-\frac{1}{a^3}+\frac{3 \ln^2a}{2 a^2}+\frac{3 \ln a}{2 a^2} \right) \right. \right.\nonumber\\ &&\left. \left.+ \sum_{n\neq 2} \left( \frac{6 a^{-2}}{n^2 (n+1)(n-1)(n-2)}+\frac{a^{-n-3}}{ n(n+1)}-\frac{3 a^{-n-2}}{ n^2}+\frac{3 a^{-n-1}}{n(n-1)}-\frac{a^{-n}}{n(n-2)} \right) + \frac{1}{6 a^5}-\frac{3}{4 a^4}+\frac{3}{2a^3}-\frac{11}{12 a^2}+\frac{\ln a}{2a^2}\right]\right. \nonumber\\ && \left.+\frac43 \left( \frac{2}{a^3}+\frac{15}{4 a^2}+\frac{\ln^3a}{a^2}+\frac{3 \ln^2a}{2 a^2}+\frac{15 \ln a}{2 a^2}-\frac{6}{a}+\frac{1}{4}\right)\right. \nonumber\\ && \left.
+\frac43\sum\frac{1}{mn} \left( \frac{6 a^{-2}}{(m+n) (m+n+1)(m+n-1)(m+n-2)}+\frac{a^{-m-n-3}}{ m+n+1}-\frac{3 a^{-m-n-2}}{m+n}+\frac{3 a^{-m-n-1}}{m+n-1}-\frac{a^{-n-m}}{m+n-2} \right)\right. \nonumber\\ && \left.+\frac83\sum_{n\neq 1,2}\frac{1}{n}\left(-\frac{12 a^{-2}}{(n-2)^2 (n-1)^2 n^2 (n+1)^2}+\frac{6 n^2 a^{-2} \ln a}{(n-2)^2 (n-1)^2 (n+1)^2}-\frac{24 n a^{-2}}{(n-2)^2 (n-1)^2 (n+1)^2} \right. \right. \nonumber\\ &&\left.\left. +\frac{36 a^{-2}}{(n-2)^2 (n-1)^2 (n+1)^2}+\frac{12 a^{-2}}{(n-2)^2 (n-1)^2 n (n+1)^2}+\frac{12 a^{-2} \ln a}{(n-2)^2 (n-1)^2 n (n+1)^2} \right. \right. \nonumber\\ &&\left.\left. -\frac{12 n a^{-2} \ln a}{(n-2)^2 (n-1)^2 (n+1)^2}-\frac{6 a^{-2} \ln a}{(n-2)^2 (n-1)^2 (n+1)^2}-\frac{a^{-n-3}}{ (n+1)^2}+\frac{3 a^{-n-2}}{ n^2}-\frac{3 a^{-n-1}}{ (n-1)^2}+\frac{a^{-n}}{(n-2)^2} \right)\right.\nonumber\\
&&\left. {-\frac{4}{27 a^5}+\frac{1}{3 a^4}+\frac{4}{a^3}-\frac{185}{27 a^2}-\frac{10 \ln ^2a}{3 a^2}+\frac{14 \ln a}{9 a^2}+\frac{8}{3 a}}
\right]\nonumber\\
&&=- \frac{\beta^2 H^2}{2^{7}\pi^4}\left[\left(\frac{145}{144}+\frac{\pi^2}{6} -\frac{8}{3} \ln \frac{2\mu}{\sqrt{e} H} +  \ln^2 \frac{2\mu}{\sqrt{e} H} \right)+\left(1-2 \ln\frac{2\mu}{\sqrt{e}H} \right)\ln a +\ln^2a \right.\nonumber\\&&\left.
+ \left(\frac{4}{9}-\frac13 \ln\frac{2\mu}{\sqrt{e}H} \right) \sum \left( \frac{(n+2)(n+3)}{a^{n+1}} -\frac{3(n^2-1)(n+2)}{na^n} +\frac{3(n+1)(n-2)}{a^{n-1}}- \frac{(n-1)(n-3)}{a^{n-2}}\right)\right. \nonumber\\ &&\left. +\frac16 \sum\frac{1}{mn}\left(\frac{(m+n)(m+n+2)(m+n+3)}{a^{m+n+1}}-  \frac{3(m+n-1)(m+n+1)(m+n+2)}{a^{m+n}}  \right.\right. \nonumber\\ && \left.\left.+ \frac{3(m+n)(m+n+1)(m+n-2)}{a^{m+n-1}} -\frac{(m+n)(m+n-1)(m+n-3)}{a^{m+n-2}}\right) -2\sum \frac{a^{-n-1}}{(n+1)^2} +\frac{5}{18a^3}-\frac{5}{4 a^2} \right]
\label{c7}
\end{eqnarray}
where we have also used,
\begin{eqnarray}
\sum_{n=3}^{\infty} \frac{(n^2-2n-1+2/n)}{n(n-1)^2(n-2)^2(n+1)^2}=\frac{\pi^2}{12} -\frac{29}{36}
\label{sum1}
\end{eqnarray}
%

%
%
The sixth integral corresponding to \ref{c3} is given by (abbreviating $\mu'=H/2e^{1/4}$),
{\footnotesize
\begin{eqnarray}
    &&\frac{i\beta^2 H^4}{2^9 \pi^6 a} \int d^4 x' a'^3\left(\frac{\ln^2 \frac{\sqrt{e} H^2\Delta x_{++}^2}{4}}{\Delta x_{++}^2}-\frac{\ln^2 \frac{\sqrt{e} H^2\Delta x_{+-}^2}{4}}{\Delta x_{+-}^2} \right)
= \frac{\beta^2}{3\times 2^8 \pi^4 a}\p_0^2 \int \frac{da'}{a'^2}\left(1-\frac{a'}{a} \right)^3\nonumber\\
&&\times \left[\left( \frac{113}{9} -\frac{2\pi^2}{3}-\frac{32}{3} \ln \frac{2\mu'}{H}+4\ln^2 \frac{2\mu'}{H}\right)+ \left(\frac{32}{3}-8 \ln \frac{2\mu'}{H} \right)\ln a' +4\ln^2a'  \right.\nonumber\\ &&\left.+\left(\frac{32}{3}-8 \ln \frac{2\mu'}{H} \right)\sum \frac{a'^n}{na^n} + 4 \sum \frac{a'^{m+n}}{mn\, a^{m+n}} +8 \ln a' \sum \frac{a'^n}{na^n} \right]\nonumber\\
&&= \frac{\beta^2 H^2}{2^7 \pi^4 a} \int da' \left(1-\frac{a'}{a} \right)
\left[\left( \frac{113}{9} -\frac{2\pi^2}{3}-\frac{32}{3} \ln \frac{2\mu'}{H}+4\ln^2 \frac{2\mu'}{H}\right)+ \left(\frac{32}{3}-8 \ln \frac{2\mu'}{H} \right)\ln a' +4\ln^2a'  \right.\nonumber\\ &&\left.+\left(\frac{32}{3}-8 \ln \frac{2\mu'}{H} \right)\sum \frac{a'^n}{na^n} + 4 \sum \frac{a'^{m+n}}{mn\, a^{m+n}} +8 \ln a' \sum \frac{a'^n}{na^n} \right]
\nonumber\\
 &&= \frac{\beta^2 H^2}{2^7 \pi^4 }\left[\frac{101}{72}-\pi^2 +\frac{1}{a}\left(\frac{2\pi^2}{3}-\frac{389}{36} \right)- \frac{1}{a^2}\left(\frac{\pi^2}{3}+\frac{55}{72} \right)+\frac73\ln a + 2\ln^2 a \right. \nonumber\\ && \left.+\sum \frac{2}{n(n+1)}\left(\frac{(7n+26)a^{-n-2}}{3(n+2)^2} +\frac{4 a^{-n-1}}{(n+1)}\right) +4 \sum \frac{1}{mn}\left(\frac{a^{-m-n-2}}{(m+n+2)} -\frac{a^{-m-n-1}}{(m+n+1)} \right) \right]
\label{c8}
\end{eqnarray}
}
Finally, the last integral equals
{\footnotesize
\begin{eqnarray}
    &&-\frac{i\beta^2 H^6}{3\times 2^{10} \pi^6 }\int d^4 x' a'^4 \left( \ln^3 \frac{\sqrt{e} H^2\Delta x_{++}^2}{4}-\ln^3 \frac{\sqrt{e} H^2\Delta x_{+-}^2}{4}\right) = \frac{\beta^2 H^2}{9\times 2^{7} \pi^4 }\int \frac{da'}{a'}\left(1-\frac{a'}{a} \right)^3\nonumber\\
    && \times \left[\left(\frac{16}{9}-\frac{\pi^2}{6}-\frac{8}{3}\ln \frac {2\mu'}{H}+  \ln^2 \frac{2\mu'}{H} \right) +  \left(\frac{8}{3} - 2 \ln \frac{2\mu'}{H}\right) \ln a'  +\ln^2 a'  + \left(\frac{8}{3} - 2 \ln \frac{2\mu'}{H}\right) \sum \frac{a'^n}{n a^n} \right. \nonumber\\ && \left. + \sum \frac{a'^{m+n}}{mn\, a^{m+n}} + 2 \ln a'   \sum \frac{a'^n}{n a^n}     \right]\nonumber\\
    &&= \frac{\beta^2 H^2}{9\times 2^{7} \pi^4 } \left[ \left(\frac{361}{144}-\frac{\pi ^2}{6} \right) \left(\frac{1}{3 a^3}-\frac{3}{2 a^2}+\frac{3}{a}+\ln a-\frac{11}{6} \right)+ \frac{19}{6}\left( -\frac{1}{9 a^3}+\frac{3}{4 a^2}-\frac{3}{a}+\frac{\ln^2a}{2}-\frac{11 \ln a}{6}+\frac{85}{36}\right) \right. \nonumber\\ &&\left.
     +\left( \frac{2}{27 a^3}-\frac{3}{4 a^2}+\frac{6}{a}+\frac{\ln^3a}{3}-\frac{11 \ln^2a}{6}+\frac{85 \ln a}{18}-\frac{575}{108}\right)\right. \nonumber\\ &&\left.
+ \frac{19}{6}  \left(\frac{\pi ^2}{6}-\frac{49}{36}+\sum\left(\frac{a^{-n-3}}{n(n+3)} -\frac{3 a^{-n-2}}{n (n+2)}+\frac{3 a^{-n-1}}{n (n+1)}-\frac{a^{-n}}{n^2}\right) \right)  \right. \nonumber\\&& \left.+\sum\frac{1}{mn} \left(\frac{6 }{(m+n) (m+n+1) (m+n+2) (m+n+3)}+\frac{a^{-m-n-3}}{ (m+n+3)}-\frac{3 a^{-m-n-2}}{ (m+n+2)}+\frac{3 a^{-m-n-1}}{ (m+n+1)}-\frac{a^{-m-n}}{m+n}  \right) \right. \nonumber\\&& \left.
-2\left(\zeta (3)+\frac{11 \pi ^2}{36}-\frac{395}{108} + \left(\frac{49}{36}-\frac{\pi ^2}{6} \right) \ln a  \right)
 \right] \nonumber\\
 &&= \frac{\beta^2 H^2}{9\times 2^{7} \pi^4 } \left[ \frac{2 \pi ^2}{9}+\frac{485}{864}-2 \zeta (3) +\left(\frac{193}{48}-\frac{\pi ^2}{2}\right)\frac{1}{a}   +\left( \frac{\pi ^2}{4}-\frac{205}{96}\right) \frac{1}{a^2} +\left( \frac{241}{432}-\frac{\pi ^2}{18}\right)\frac{1}{a^3} + \left(\frac{\pi ^2}{6}-\frac{187}{144} \right)\ln a  \right. \nonumber\\&& \left. -\frac14\ln^2 a +\frac13 \ln^3 a +\frac{19}{6}\sum\left(\frac{a^{-n-3}}{n(n+3)} -\frac{3 a^{-n-2}}{n (n+2)}+\frac{3 a^{-n-1}}{n (n+1)}-\frac{a^{-n}}{n^2}\right)  \right. \nonumber\\ && \left. 
+\sum\frac{1}{mn} \left(\frac{6 }{(m+n) (m+n+1) (m+n+2) (m+n+3)}+\frac{a^{-m-n-3}}{ (m+n+3)}-\frac{3 a^{-m-n-2}}{ (m+n+2)}+\frac{3 a^{-m-n-1}}{ (m+n+1)}-\frac{a^{-m-n}}{m+n}  \right)  \right] \nonumber\\
\label{c9}
\end{eqnarray}
}
Combining now \ref{c4}, \ref{c5}, \ref{c7}, \ref{c8} and \ref{c9}, and after a little bit of rearrangements, \ref{c10''} in the main text follows.


\begin{thebibliography}{99} 


\bibitem{Weinberg-Book}
S.~Weinberg, {\it Cosmology}, Oxford Univ. Press (2009).

\bibitem{Tsamis:2005hd}
N.~C.~Tsamis and R.~P.~Woodard,
{\it Stochastic quantum gravitational inflation},
Nucl.~Phys.~B\textbf{724}, 295 (2005)
[arXiv:gr-qc/0505115 [gr-qc]].

\bibitem{Birrell:1982ix} 
N.~D.~Birrell and P.~C.~W.~Davies,
{\it Quantum Fields in Curved Space}, Cambridge Univ. Press (1982).



\bibitem{Floratos:1987ek}
E.~G.~Floratos, J.~Iliopoulos and T.~N.~Tomaras,
{\it Tree Level Scattering Amplitudes in De Sitter Space Diverge},
Phys.~Lett.~B\textbf{197}, 373 (1987).

\bibitem{Tanaka:2013caa}
T.~Tanaka and Y.~Urakawa,
{\it Loops in inflationary correlation functions},
Class. Quant. Grav.\textbf{30}, 233001 (2013)
[arXiv:1306.4461 [hep-th]].



\bibitem{Tsamis:1992sx}
N.~C.~Tsamis and R.~P.~Woodard,
{\it Relaxing the cosmological constant},
Phys.~Lett.~B\textbf{301}, 351 (1993)



\bibitem{Ringeval:2010hf}
C.~Ringeval, T.~Suyama, T.~Takahashi, M.~Yamaguchi and S.~Yokoyama,
{\it Dark energy from primordial inflationary quantum fluctuations},
Phys. Rev. Lett.\textbf{105}, 121301 (2010)
[arXiv:1006.0368 [astro-ph.CO]].

\bibitem{Dadhich:2011gx}
N.~Dadhich,
{\it On the measure of spacetime and gravity},
Int. J. Mod. Phys. D\textbf{20}, 2739-2747 (2011)
[arXiv:1105.3396 [gr-qc]].

\bibitem{Padmanabhan:2013hqa}
T.~Padmanabhan and H.~Padmanabhan,
{\it CosMIn: The Solution to the Cosmological Constant Problem},
Int. J. Mod. Phys. D\textbf{22}, 1342001 (2013)
[arXiv:1302.3226 [astro-ph.CO]].

\bibitem{Alberte:2016izw}
L.~Alberte, P.~Creminelli, A.~Khmelnitsky, D.~Pirtskhalava and E.~Trincherini,
{\it Relaxing the Cosmological Constant: a Proof of Concept},
JHEP\textbf{12}, 022 (2016)
[arXiv:1608.05715 [hep-th]].


\bibitem{Appleby:2018yci}
S.~Appleby and E.~V.~Linder,
{\it The Well-Tempered Cosmological Constant},
JCAP\textbf{07}, 034 (2018)
[arXiv:1805.00470 [gr-qc]].

\bibitem{Starobinsky}
A.~A.~Starobinsky, Lect. Notes Phys.{\bf246}, 107 (1986) doi:10.1007/3-540-16452-9\_6

\bibitem{Starobinsky:1994bd}
A.~A.~Starobinsky and J.~Yokoyama,
{\it Equilibrium state of a selfinteracting scalar field in the De Sitter background},
Phys.~Rev.~D\textbf{50}, 6357 (1994)
[arXiv:astro-ph/9407016 [astro-ph]].

\bibitem{Finelli:2008zg}
F.~Finelli, G.~Marozzi, A.~A.~Starobinsky, G.~P.~Vacca and G.~Venturi,
{\it Generation of fluctuations during inflation: Comparison of stochastic and field-theoretic approaches},
Phys.~Rev.~D\textbf{79}, 044007 (2009)
[arXiv:0808.1786 [hep-th]].


\bibitem{Vennin:2015hra}
V.~Vennin and A.~A.~Starobinsky,
{\it Correlation Functions in Stochastic Inflation},
Eur.~Phys.~J.~C\textbf{75}, 413 (2015)
[arXiv:1506.04732 [hep-th]].

\bibitem{Markkanen:2019kpv}
T.~Markkanen, A.~Rajantie, S.~Stopyra and T.~Tenkanen,
{\it Scalar correlation functions in de Sitter space from the stochastic spectral expansion},
JCAP\textbf{08}, 001 (2019)
[arXiv:1904.11917 [gr-qc]].

\bibitem{Markkanen:2020bfc}
T.~Markkanen and A.~Rajantie,
{\it Scalar correlation functions for a double-well potential in de Sitter space},
JCAP\textbf{03}, 049 (2020)
[arXiv:2001.04494 [gr-qc]].

\bibitem{Moreau:2018ena}
G.~Moreau and J.~Serreau,
{\it Backreaction of superhorizon scalar field fluctuations on a de Sitter geometry: A renormalization group perspective},
Phys.~Rev.~D\textbf{99}, no.2, 025011 (2019)
[arXiv:1809.03969 [hep-th]].

\bibitem{Moreau:2018lmz}
G.~Moreau and J.~Serreau,
{\it Stability of de Sitter spacetime against infrared quantum scalar field fluctuations},
Phys.~Rev.~Lett.\textbf{122}, no.1, 011302 (2019)
[arXiv:1808.00338 [hep-th]].

\bibitem{Gautier:2015pca}
F.~Gautier and J.~Serreau,
{\it Scalar field correlator in de Sitter space at next-to-leading order in a 1/N expansion},
Phys.~Rev.~D\textbf{92}, no.10, 105035 (2015)
[arXiv:1509.05546 [hep-th]].

\bibitem{Serreau:2013eoa}
J.~Serreau,
{\it Renormalization group flow and symmetry restoration in de Sitter space},
Phys.~Lett.~B\textbf{730}, 271 (2014)
[arXiv:1306.3846 [hep-th]].

\bibitem{Serreau:2013koa}
J.~Serreau,
{\it Nonperturbative infrared enhancement of nonGaussian correlators in de Sitter space},
Phys.~Lett.~B\textbf{728}, 380 (2014)
[arXiv:1302.6365 [hep-th]].

\bibitem{Serreau:2013psa}
J.~Serreau and R.~Parentani,
{\it Nonperturbative resummation of de Sitter infrared logarithms in the large-N limit},
Phys.~Rev.~D\textbf{87}, 085012 (2013)
[arXiv:1302.3262 [hep-th]].

\bibitem{Ferreira:2017ogo}
R.~Z.~Ferreira, M.~Sandora and M.~S.~Sloth,
{\it Patient Observers and Non-perturbative Infrared Dynamics in Inflation},
JCAP\textbf{02}, 055 (2018)
[arXiv:1703.10162 [hep-th]].


\bibitem{Burgess:2009bs}
C.~P.~Burgess, L.~Leblond, R.~Holman and S.~Shandera,
{\it Super-Hubble de Sitter Fluctuations and the Dynamical RG},
JCAP\textbf{03}, 033 (2010)
[arXiv:0912.1608 [hep-th]].

\bibitem{Burgess:2015ajz}
C.~P.~Burgess, R.~Holman and G.~Tasinato,
{\it Open EFTs, IR effects $\&$ late-time resummations: systematic corrections in stochastic inflation},
JHEP\textbf{01}, 153 (2016)
[arXiv:1512.00169 [gr-qc]].

\bibitem{Youssef:2013by}
A.~Youssef and D.~Kreimer,
{\it Resummation of infrared logarithms in de Sitter space via Dyson-Schwinger equations: the ladder-rainbow approximation},
Phys. Rev. D\textbf{89}, 124021 (2014)
[arXiv:1301.3205 [gr-qc]].

\bibitem{Baumgart:2019clc} 
  M.~Baumgart and R.~Sundrum,
{\it De Sitter Diagrammar and the Resummation of Time},
  arXiv:1912.09502.


\bibitem{Kamenshchik:2020yyn}
A.~Y.~Kamenshchik and T.~Vardanyan,
{\it Renormalization group inspired autonomous equations for secular effects in de Sitter space},
Phys. Rev. D\textbf{102}, no.6, 065010 (2020)
[arXiv:2005.02504 [hep-th]].


\bibitem{Kamenshchik:2021tjh}
A.~Y.~Kamenshchik, A.~A.~Starobinsky and T.~Vardanyan,
{\it Massive scalar field in de Sitter spacetime: a two-loop calculation and a comparison with the stochastic approach},
Eur.~Phys.~J.~C\textbf{82}, no.4, 345 (2022)
[arXiv:2109.05625 [gr-qc]].


\bibitem{Moss:2016uix}
I.~Moss and G.~Rigopoulos,
{\it Effective long wavelength scalar dynamics in de Sitter},
JCAP\textbf{05}, 009 (2017)
[arXiv:1611.07589 [gr-qc]].

\bibitem{Chu:2015ila}
C.~S.~Chu and Y.~Koyama,
{\it Dilaton, Screening of the Cosmological Constant and IR-Driven Inflation},
JHEP\textbf{09}, 024 (2015)
[arXiv:1506.02848 [hep-th]].

\bibitem{Kitamoto:2011yx}
H.~Kitamoto and Y.~Kitazawa,
{\it Infra-red effects of Non-linear sigma model in de Sitter space},
Phys.~Rev.~D\textbf{85}, 044062 (2012)
[arXiv:1109.4892 [hep-th]].

\bibitem{Kitamoto:2018dek}
H.~Kitamoto,
{\it Infrared resummation for derivative interactions in de Sitter space},
Phys.~Rev.~D\textbf{100}, no.2, 025020 (2019)
[arXiv:1811.01830 [hep-th]].


\bibitem{Miao:2021gic} 
S.~P.~Miao, N.~C.~Tsamis and R.~P.~Woodard,
{\it Summing Inflationary Logarithms in Nonlinear Sigma Models},
[arXiv:2110.08715 [gr-qc]].

\bibitem{Woodard:2014jba} 
  R.~P.~Woodard,
  {\it Perturbative Quantum Gravity Comes of Age},
  Int.\ J.\ Mod.\ Phys.\ D{\bf 23}, no. 09, 1430020 (2014)
  [arXiv:1407.4748 [gr-qc]].



 
\bibitem{Onemli:2002hr}
V.~K.~Onemli and R.~P.~Woodard,
{\it Superacceleration from massless, minimally coupled $\phi^4$},
Class.~Quant.~Grav.\textbf{19}, 4607 (2002)
[arXiv:gr-qc/0204065 [gr-qc]].
  
 \bibitem{Brunier:2004sb}
T.~Brunier, V.~K.~Onemli and R.~P.~Woodard,
{\it Two loop scalar self-mass during inflation},
Class. Quant. Grav.\textbf{22}, 59 (2005)
[arXiv:gr-qc/0408080 [gr-qc]].

\bibitem{Kahya:2009sz}
E.~O.~Kahya, V.~K.~Onemli and R.~P.~Woodard,
{\it A Completely Regular Quantum Stress Tensor with $w \ensuremath{<} -1$},
Phys. Rev. D\textbf{81}, 023508 (2010)
[arXiv:0904.4811 [gr-qc]].

\bibitem{Boyanovsky:2012qs}
D.~Boyanovsky,
{\it Condensates and quasiparticles in inflationary cosmology: mass generation and decay widths},
Phys. Rev. D\textbf{85}, 123525 (2012)
[arXiv:1203.3903 [hep-ph]].



\bibitem{Onemli:2015pma}
V.~K.~Onemli,
{\it Vacuum Fluctuations of a Scalar Field during Inflation: Quantum versus Stochastic Analysis},
Phys.~Rev.~D\textbf{91}, 103537 (2015)
[arXiv:1501.05852 [gr-qc]].

\bibitem{Karakaya:2017evp}
G.~Karakaya and V.~K.~Onemli,
{\it Quantum effects of mass on scalar field correlations, power spectrum, and fluctuations during inflation},
Phys.~Rev.~D\textbf{97}, no.12, 123531 (2018)
[arXiv:1710.06768 [gr-qc]].


  
    \bibitem{Ali:2020gij}
M.~S.~Ali, S.~Bhattacharya and K.~Lochan,
{\it Unruh-DeWitt detector responses for complex scalar fields in de Sitter spacetime},
JHEP\textbf{03}, 220 (2021)
[arXiv:2003.11046 [hep-th]].

  
   
   \bibitem{Prokopec:2003tm}
  T.~Prokopec and E.~Puchwein,
{\it Photon mass generation during inflation: de Sitter invariant case},
  JCAP{\bf 0404}, 007 (2004)
  [astro-ph/0312274].
  


\bibitem{Miao:2006pn}
  S.~P.~Miao and R.~P.~Woodard,
{\it Leading log solution for inflationary Yukawa},
  Phys.\ Rev.\ D{\bf 74}, 044019 (2006)
  [gr-qc/0602110].
  
  \bibitem{Prokopec:2007ak}
T.~Prokopec, N.~C.~Tsamis and R.~P.~Woodard,
{\it Stochastic Inflationary Scalar Electrodynamics},
Annals Phys.\textbf{323}, 1324 (2008)
[arXiv:0707.0847 [gr-qc]].

\bibitem{Liao:2018sci}
J.~H.~Liao, S.~P.~Miao and R.~P.~Woodard,
{\it Cosmological Coleman-Weinberg Potentials and Inflation},
Phys. Rev. D\textbf{99}, no.10, 103522 (2019)
[arXiv:1806.02533 [gr-qc]].
  
  \bibitem{Miao:2020zeh}
S.~P.~Miao, L.~Tan and R.~P.~Woodard,
{\it Bose\textendash{}Fermi cancellation of cosmological Coleman\textendash{}Weinberg potentials},
Class. Quant. Grav.\textbf{37}, no.16, 165007 (2020)
[arXiv:2003.03752 [gr-qc]].


\bibitem{Glavan:2019uni}
D.~Glavan and G.~Rigopoulos,
{\it One-loop electromagnetic correlators of SQED in power-law inflation},
JCAP\textbf{02}, 021 (2021)
[arXiv:1909.11741 [gr-qc]].
  
  
   


\bibitem{Giddings:2010nc}
S.~B.~Giddings and M.~S.~Sloth,
{\it Semiclassical relations and IR effects in de Sitter and slow-roll space-times},
JCAP\textbf{01}, 023 (2011)
[arXiv:1005.1056 [hep-th]].


\bibitem{Leonard:2014zua}
K.~E.~Leonard, S.~Park, T.~Prokopec and R.~P.~Woodard,
{\it Representing the Graviton Self-Energy on de Sitter Background},
Phys.~Rev.~D\textbf{90}, no.2, 024032 (2014)
[arXiv:1403.0896 [gr-qc]].

\bibitem{Park:2015kua}
S.~Park, T.~Prokopec and R.~P.~Woodard,
{\it Quantum Scalar Corrections to the Gravitational Potentials on de Sitter Background},
JHEP\textbf{01}, 074 (2016)
[arXiv:1510.03352 [gr-qc]].


\bibitem{Frob:2016fcr}
M.~B.~Fr\"ob and E.~Verdaguer,
{\it Quantum corrections to the gravitational potentials of a point source due to conformal fields in de Sitter},
JCAP \textbf{03}, 015 (2016)
[arXiv:1601.03561 [hep-th]].


\bibitem{Frob:2017smg}
M.~B.~Fr\"{o}b and E.~Verdaguer,
{\it Quantum corrections for spinning particles in de Sitter},
JCAP\textbf{04}, 022 (2017)
[arXiv:1701.06576 [hep-th]].

\bibitem{Frob:2017lnt}
M.~B.~Fr\"ob,
{\it Gauge-invariant quantum gravitational corrections to correlation functions},
Class. Quant. Grav.\textbf{35}, no.5, 055006 (2018)
[arXiv:1710.00839 [gr-qc]].


\bibitem{Boran:2017fsx}
S.~Boran, E.~O.~Kahya and S.~Park,
{\it Quantum gravity corrections to the conformally coupled scalar self-mass-squared on de Sitter background. II. Kinetic conformal cross terms},
Phys. Rev. D\textbf{96}, no.2, 025001 (2017)
[arXiv:1704.05880 [gr-qc]].

\bibitem{Boran:2017cfj}
S.~Boran, E.~O.~Kahya and S.~Park,
{\it One loop corrected conformally coupled scalar mode equations during inflation},
Phys. Rev. D\textbf{96}, no.10, 105003 (2017)
[erratum: Phys. Rev. D\textbf{98}, no.2, 029903 (2018)]
[arXiv:1708.01831 [gr-qc]].

\bibitem{Miao:2018bol}
S.~P.~Miao, T.~Prokopec and R.~P.~Woodard,
{\it Scalar enhancement of the photon electric field by the tail of the graviton propagator},
Phys. Rev. D\textbf{98}, no.2, 025022 (2018)
[arXiv:1806.00742 [gr-qc]].

\bibitem{Ferrero:2021lhd}
R.~Ferrero and C.~Ripken,
{\it De Sitter scattering amplitudes in the Born approximation},
[arXiv:2112.03766 [hep-th]].

\bibitem{Tan:2021ibs}
L.~Tan, N.~C.~Tsamis and R.~P.~Woodard,
{\it Graviton self-energy from gravitons in cosmology},
Class.~Quant.~Grav.\textbf{38}, no.14, 145024 (2021)
[arXiv:2103.08547 [gr-qc]].

\bibitem{Glavan:2021adm}
D.~Glavan, S.~P.~Miao, T.~Prokopec and R.~P.~Woodard,
{\it Large Logarithms from Quantum Gravitational Corrections to a Massless, Minimally Coupled Scalar on de Sitter},
[arXiv:2112.00959 [gr-qc]].


\bibitem{Akhmedov:2013xka}
E.~T.~Akhmedov, F.~K.~Popov and V.~M.~Slepukhin,
{\it Infrared dynamics of the massive \ensuremath{\phi}4 theory on de Sitter space},
Phys.~Rev.~D\textbf{88}, 024021 (2013)
[arXiv:1303.1068 [hep-th]].


\bibitem{Akhmedov:2014doa}
E.~T.~Akhmedov and F.~K.~Popov,
{\it A few more comments on secularly growing loop corrections in strong electric fields},
JHEP\textbf{09}, 085 (2015)
[arXiv:1412.1554 [hep-th]].

\bibitem{Akhmedov:2015xwa}
E.~T.~Akhmedov, H.~Godazgar and F.~K.~Popov,
{\it Hawking radiation and secularly growing loop corrections},
Phys. Rev. D\textbf{93}, no.2, 024029 (2016)
[arXiv:1508.07500 [hep-th]].

\bibitem{Akhmedov:2019cfd}
E.~T.~Akhmedov, U.~Moschella and F.~K.~Popov,
{\it Characters of different secular effects in various patches of de Sitter space},
Phys. Rev. D\textbf{99}, no.8, 086009 (2019)
[arXiv:1901.07293 [hep-th]].

\bibitem{Kaplanek:2020iay}
G.~Kaplanek and C.~P.~Burgess,
{\it Qubits on the Horizon: Decoherence and Thermalization near Black Holes},
JHEP\textbf{01}, 098 (2021)
[arXiv:2007.05984 [hep-th]].

\bibitem{Burgess:2018sou}
C.~P.~Burgess, J.~Hainge, G.~Kaplanek and M.~Rummel,
{\it Failure of Perturbation Theory Near Horizons: the Rindler Example},
JHEP\textbf{10}, 122 (2018)
[arXiv:1806.11415 [hep-th]].


\bibitem{Hu:2018nxy}
B.~L.~Hu,
{\it Infrared Behavior of Quantum Fields in Inflationary Cosmology -- Issues and Approaches: an overview},
[arXiv:1812.11851 [gr-qc]].

\bibitem{Martins:2020oxv}
J.~S.~Martins, O.~F.~Piattella, I.~L.~Shapiro and A.~A.~Starobinsky,
{\it Inflation with sterile scalar coupled to massive fermions and to gravity},
[arXiv:2010.14639 [hep-th]].




\bibitem{Calzetta:2008iqa}
E.~A.~Calzetta and B.~L.~B.~Hu,
{\it Nonequilibrium Quantum Field Theory},
Cambridge Univ. Press (2008).


\bibitem{Chou}
K.~Chou and Z.~Su and B.~Hao and L.~Yu, {\it Equilibrium and nonequilibrium formalisms made unified}, Phys.~Rep.{\bf118}, 1 (1985).

\bibitem{Calzetta1}
E.~Calzetta and B.~L.~Hu, {\it Closed Time-Path Functional Formalism in Curved Spacetime: Application to Cosmological Back-Reaction Problems}, 
Phys.~Rev.~D{\bf35}, 495 (1987).

\bibitem{Calzetta2}
E.~Calzetta and B.~L.~Hu, {\it Nonequilibrium quantum fields: Closed-time-path effective action, Wigner function, and Boltzmann equation}, Phys.~Rev.~D{\bf37}, 2878 (1988).

\bibitem{Weinberg}
S.~Weinberg, {\it Quantum contributions to cosmological correlations}, Phys.~Rev.~D{\bf 72}, 043514 (2005).

\bibitem{Adshead:2009cb}
P.~Adshead, R.~Easther and E.~A.~Lim,
{\it The `in-in' Formalism and Cosmological Perturbations},
Phys. Rev. D\textbf{80}, 083521 (2009)
[arXiv:0904.4207 [hep-th]].





\bibitem{GR}
 M.~Abramowitz, I.~A.~Stegun, {\it Handbook of Mathematical Functions With Formulas, Graphs, and Mathematical Tables}, as available at www.convertit.com/Go/ConvertIt/Reference/AMS55.ASP 266.

\bibitem{Enqvist:2017kzh}
K.~Enqvist, R.~J.~Hardwick, T.~Tenkanen, V.~Vennin and D.~Wands,
{\it A novel way to determine the scale of inflation},
JCAP\textbf{02}, 006 (2018)
[arXiv:1711.07344 [astro-ph.CO]].






  \end{thebibliography}
\end{document}